\documentclass[smallextended]{svjour3}       

\usepackage[table,xcdraw,dvipsnames]{xcolor} 
\usepackage[utf8]{inputenc} 
\usepackage[T1]{fontenc}
\usepackage{graphicx}
\usepackage{glossaries}
\usepackage{soul} 
\usepackage[bookmarksdepth=2]{hyperref} 
\usepackage{bbding}
\usepackage{float} 
\usepackage[sort&compress,square,comma,numbers]{natbib}
\usepackage[many]{tcolorbox} 
\usepackage{enumitem} 
\usepackage{ifthen} 
\usepackage{amssymb} 
\usepackage{booktabs}
\usepackage{graphicx}
\usepackage{makecell}
\usepackage{amsmath}
\usepackage[export]{adjustbox}
\usepackage{ragged2e} 
\usepackage{relsize} 
\usepackage{multirow} 
\usepackage{pbox} 
\usepackage{balance} 
\usepackage{moresize} 
\usepackage{xspace}

\usepackage{placeins} 

\newtcolorbox{mybox}[2][]{
top=0.15in,left=4pt,right=4pt,bottom=4pt,
fonttitle=\bfseries,
colbacktitle=gray,
colback=gray!5,
colframe=gray!40!black,
enhanced,
attach boxed title to top left={xshift=1.5em,yshift=-\tcboxedtitleheight/2},
boxed title style={size=small},
drop shadow={black!50!white},
title=#2,#1}

\newcommand{\countobservations}{
    \def \countobservations{1}
}
\newcounter{observation}
\newenvironment{observation}[1]
{   \refstepcounter{observation}
    \noindent \textbf{Observation~\theobservation) \textit{#1}}
}

\countobservations

\newcommand{\countimplications}{
    \def \countimplications{1}
}
\newcounter{implication}
\countimplications

\newboolean{showcomments}
\setboolean{showcomments}{true}

\ifthenelse{\boolean{showcomments}}
{\newcommand{\nbc}[3]{
 {\colorbox{#3}{\bfseries\sffamily\scriptsize\textcolor{white}{#1}}}
 {\textcolor{#3}{\sf\small$\blacktriangleright$\textit{#2}$\blacktriangleleft$}}
 }
}
{\newcommand{\nbc}[3]{}
 }

\newcommand{\smalltt}[1]{\ifmmode{\mbox{\smaller\texttt{#1}}}\else{\smaller\tt #1}\fi}

\newcolumntype{L}[1]{>{\raggedright\let\newline\\\arraybackslash\hspace{0pt}}m{#1}}
\newcolumntype{C}[1]{>{\centering\let\newline\\\arraybackslash\hspace{0pt}}m{#1}}
\newcolumntype{R}[1]{>{\raggedleft\let\newline\\\arraybackslash\hspace{0pt}}m{#1}}

\usepackage{soul}
\usepackage{moresize}
\usepackage{listings}
\usepackage{subcaption}

\begin{document}

	\title{Assessing and Improving the Representativeness of Code Generation Benchmarks Using Knowledge Units (KUs) of Programming Languages – An Empirical Study}
	
	\titlerunning{Studying the Representativeness of Code Generation Benchmarks with KUs}
	
	\author{Md Ahasanuzzaman \and Bram Adams \and Emad Fallahzadeh \and Gustavo A. Oliva  \and Ahmed E. Hassan}

	\institute{
		\Envelope \space Md Ahasanuzzaman, Bram Adams, Emad Fallahzadeh, Gustavo A. Oliva and Ahmed E. Hassan \at
		School of Computing \\
		Queen's University, Kingston, Ontario, Canada\\    
		\email{\{md.ahasanuzzaman, bram.adams, emad.fallahzadeh, gustavo.oliva, hassan\}@queensu.ca}
	}

	\date{Received: date / Accepted: date}

	\maketitle

	\begin{abstract}
		\justifying{
Large Language Models (LLMs) such as GPT-4, Calude, LLaMA and StarCoder have achieved notable progress in code generation tasks, with benchmark datasets like HumanEval and MBPP serving as the primary means of evaluating their code-generation performance. Effective code generation requires models to understand and correctly apply the underlying programming language concepts (e.g., concurrency). If the language concepts exercised in benchmark tasks are not representative of those employed in real-world software projects, evaluation results based on these benchmarks may yield incomplete and less reliable conclusions. Yet, despite the importance of representativeness, no prior work has systematically examined whether benchmarks reflect the language concepts that developers employ in real-world software projects.


To fill this gap, this paper presents the first systematic study of the representativeness of benchmarks in terms of language concepts, using Knowledge Units (KUs) of programming languages as an analytical lens for understanding those concepts. A KU is a cohesive set of key capabilities provided by the constructs (e.g., try/except and await) and APIs (e.g., Exception and threading.Thread.) of a given programming language. We analyze two widely used code generation benchmarks in Python - HumanEval and MBPP - alongside 30 real-world Python projects to investigate KU coverage. Our findings reveal that only half of the identified 20 Python KUs are represented in each benchmark, while real-world projects employ all of them to various degrees. Although the KU distributions in real-world software projects are not uniform -- since certain KUs (e.g., variables and control flow) naturally occur more frequently -- their distributions are substantially balanced and reflect realistic usage patterns. In contrast, benchmark tasks exhibit a skewed KU distribution, with a small subset of KUs dominating most tasks. 

To address this misalignment, we develop a prompt-based LLM framework that enhances these benchmarks with newly synthesized tasks to rebalance their KU distributions and align them with the KU distribution observed in real-world projects. Using this framework, we generate 440 synthesized tasks and augment the original benchmarks by combining these new tasks with their existing task sets. Our results present that the augmented benchmarks substantially improve KU coverage and more closely align with real-world KU distributions, resulting in an improvement of over 60\% in distributional alignment compared to the original benchmarks. Evaluating state-of-the-art LLMs on these augmented benchmarks reveals a consistent and statistically significant performance drop (12.54–44.82\% with a large effect size), highlighting that the original benchmarks were over-estimating the LLMs' performance on code generation tasks due to their bias on only few KUs instead of the real KU distribution.

Our findings provide valuable insights for researchers and practitioners aiming to design more comprehensive benchmarks, to evaluate an LLM's performance more effectively, and to better understand model capabilities through the lens of KUs, ultimately guiding the improvement of LLMs.

}

	\end{abstract}

    \newcommand\PrelimStudy{Preliminary Study Title}
	
    \newcommand\RQOne{How comprehensively do existing LLM benchmark datasets cover language concepts compared to real-world software projects?}

    \newcommand\RQOneSubOne{How well do LLM benchmark datasets represent the breadth of KUs found in real-world software projects?}

    \newcommand\RQOneSubTwo{To what extent are KUs evenly distributed across tasks in existing benchmark datasets compared to real-world software projects?}

	\newcommand\RQOneSubThree{How do the characteristic of clusters formed using KUs differ between benchmark datasets and real-world software projects?}

    \newcommand\RQTwo{To what extent can synthesized tasks restore the representativeness of language concepts in existing benchmarks?}

	  \newcommand\RQTowSubOne{To what extent does augmenting existing benchmarks with synthesized KU-based tasks improve their code-concept representativeness with respect to real-world projects?}

    \newcommand\RQTowSubTwo{How well do LLMs perform in generating code for solving tasks across different KUs?}

    \section{Introduction}
\label{sec:Intro}

Large Language Models (LLMs) such as GPT-4~\citep{achiam2023gpt}, Claude~\citep{anthropic2023claude}, LLaMA~\citep{touvron2023llama}, and StarCoder~\citep{li2023starcoder} have shown remarkable progress in code generation, offering the potential to assist developers in a wide range of software engineering tasks~\citep{bistarelli2025usage,wang2024surveyse}. Their effectiveness in generating correct and reliable code depends on how well they can understand and apply the \textit{language concepts} of a programming language -- concepts such as data types, control flow (conditions and loops), exception handling, I/O operations, and concurrency mechanisms~\citep{svyatkovskiy2020intellicode, ren2023misuse, beger2025coconut}. With recent advances in model architecture, scale, and training data, LLMs have shown notable improvements in capturing these underlying language concepts and producing code that aligns more closely with the requirements of real-world programming tasks for software development~\citep{jiang2024survey}.

Such performance improvements of LLMs in code generation are typically obtained by evaluating the LLMs using benchmark datasets such as HumanEval~\citep{chen2021evaluating}, MBPP~\citep{austin2021program}, and BigCodeBench~\citep{zhuo2024bigcodebench}. These benchmarks provide programming tasks for a given programming language, along with input–output test cases that validate the correctness of the generated code, allowing reproducible evaluation and fair comparison of code generation performance across different models~\citep{jiang2024survey}. Benchmarks therefore play a central role in measuring progress in code generation and guiding the development of newer and more capable models.

However, the usefulness of these benchmarks is closely tied to how representative the benchmark tasks are for real-world tasks, for instance in terms of their goal, their use of programming language concepts and APIs, etc. Given that non-representative datasets undermine generalizability and distort conclusions~\citep{baltes2022sampling, nagappan2013diversity}, evaluations based on these benchmarks may provide an unreliable assessment of an LLM’s true code-generation capabilities for solving real-world programming tasks. Drawing on principles of distributive fairness~\citep{german2018my} -- where outcomes need not be equal but should be appropriate relative to real usage, real-world software projects found for instance on GitHub can be treated as a reference for what constitutes a ``correct'' distribution of language concepts. 

While representativeness has been examined in several areas of software engineering -- such as software fault injections~\citep{natella2010representativeness,natella2012fault}, repository mining and dataset sampling~\citep{baltes2022sampling,gorostidi2024creation}, and the recruitment of developers for empirical studies~\citep{rainer2022recruiting, kontio2008focus} -- these investigations do not address the specific conceptual demands of code-generation tasks. In particular, no prior work has systematically examined whether code-generation benchmarks reflect the language concepts that developers employ in real-world software projects. The closer the distribution of programming language concepts in benchmarks matches that of real-world projects, the more reliable and representative the benchmarks would become for evaluating the capabilities of LLMs. As a result, it remains unclear whether benchmarks offer an approximation of code-concept usage that reflects real-world projects, or whether benchmarks introduce conceptual biases that may influence the evaluation of LLMs' code generation capabilities. 

To address this gap, this paper introduces a systematic framework for analyzing the representativeness of code-generation benchmarks in terms of the language concepts they require for their resolution. We model these language concepts using Knowledge Units (KUs), where each KU is a cohesive set of key capabilities that are offered by one or more building blocks of a programming language~\citep{ahasanuzzaman2024using,ahasanuzzaman2025predictingLTC,ahasanuzzaman2024predictingDefect}. For example, the Concurrency KU in Python includes capabilities related to concurrent programming concepts such as creating and managing threads with the \texttt{threading} module, executing tasks in parallel using the \texttt{concurrent.futures} module, or enable non-blocking concurrency using \texttt{asynchronous} co-routines. A detailed discussion of KUs is provided in Section~\ref{sec:Knowledge_Unit}.

As the first study of the representativeness of benchmarks through language concepts, we focus on the two most widely used Python code-generation benchmarks -- HumanEval~\citep{chen2021evaluating} and MBPP~\citep{austin2021program}, and analyze them through the lens of KUs. To understand KU usage in practice, we also study 30 real-world Python projects selected from a widely adopted curated list provided by~\citet{engineered_project_github_EMSE_2017}. Using our LLM-based KU detector, we systematically compare the KU distributions of benchmark tasks with those of real-world projects to examine the coverage gaps between them. To bridge this coverage gap, we develop an automatic approach for generating KU-based tasks that augments existing benchmarks and improves their KU coverage to better align with the KU distributions of real-world projects, thereby enhancing the reliability of benchmark evaluation results. Our KUs can be seen as a lens that can help understand where models excel
and where they struggle, providing valuable insight into their strengths and
weaknesses. This understanding is crucial to guide improvements and advance
the development of more capable LLMs. In this paper, we address the following two research questions:

\smallskip \noindent \textbf{RQ1: \RQOne} 
To understand how comprehensively benchmarks cover KUs compared to real-world software projects, we first compute the coverage of every KU for benchmarks and real-world projects. We define the coverage of a KU \textit{K} as the ratio of the total instances of \textit{K} in a dataset to the total instances of all KUs in that dataset. We then compare the KU coverage in the benchmarks with real-world projects to assess how well the benchmarks reflect real-world KU distributions. Finally, we analyze the distributional balance
of KUs using the Lorenz curve to determine whether benchmarks emphasize certain KUs disproportionately compared to real-world projects. 

\sloppy\smallskip\noindent\textit{
\textbf{Results:} Neither HumanEval nor MBPP provides comprehensive coverage of the key Python KUs that are commonly used in real-world projects. Only 50\% of the 20 identified KUs appear in either benchmark (in fact, both benchmarks include the same set of KUs), while real-world projects use all KUs to various degrees. Many of these missing KUs are fundamental to Python programming, including Object-Oriented Programming, Database, Concurrency, and Networking KUs. Furthermore, the KU coverage observed in these benchmarks is not representative of real-world KU usage: many KUs that frequently appear in real-world projects are either severely underrepresented or entirely absent in the benchmarks. The distribution of KUs in benchmarks is also skewed, with a small number of KUs dominating most tasks, while real-world software projects exhibit a far more balanced and comprehensive distribution. Taken together, these disparities show that the KU distributions of current benchmarks lack representativeness and distributive fairness relative to real-world software projects.}

\smallskip \noindent \textbf{RQ2: \RQTwo} Our findings in RQ1 show that HumanEval and MBPP benchmark datasets do not comprehensively cover the broader spectrum of Python KUs and that KU distributions are neither representative nor distributively fair compared to real-world software projects. Many fundamental and important Python KUs are either missing or underrepresented in these benchmarks,  limiting their ability to comprehensively evaluate the capabilities of LLMs for code generation tasks. 

To address this gap, we develop a prompt-based LLM approach that uses real-world source code containing the target KU \textit{K} as a contextual input and automatically generates a task description, a corresponding solution, and test cases to evaluate the correctness of the generated solution for that task involving with the target KU \textit{K}. Using our approach, we generate 440 synthesized tasks (on average 40 tasks per KU) for the 11 KUs that are missing or underrepresented in HumanEval and MBPP. We then combine the newly generated KU-based synthesized tasks with the HumanEval and the MBPP benchmark datasets and generate augmented benchmarks: Augmented-HumanEval and Augmented-MBPP. 

To assess how well these augmented benchmarks align with real-world KU distributions, we use the Jensen–Shannon Distance (JSDistance)~\citep{endres2003new, lin2002divergence}, a symmetric and bounded metric [0,1] that quantifies the similarity between probability distributions. A lower JSDistance value (closer to 0) indicates greater similarity between the KU coverage of a benchmark and real-world projects, whereas a higher value (closer to 1) indicates greater dissimilarity. To further examine how LLMs perform on more representative benchmarks, we evaluate seven popular LLMs on the augmented datasets. We measure the accuracy of a model using the pass@k metric, a widely adopted approach for evaluating code generation performance. In this study, we select $k = 1$, $3$, and $5$ and compute the corresponding pass@1, pass@3, and pass@5 scores for every model. 

\sloppy \smallskip \noindent \textit{\textbf{Results:} 
Combining synthesized KU-based tasks with raw benchmarks improves the overall KU coverage in augmented benchmarks -- resulting in a distribution that more closely reflects that of real-world software projects. Specifically, the JSDistance between benchmark and real-world projects' KU distributions, decreases from 0.335 to 0.118 for Augmented-HumanEval and from 0.319 to 0.122 for Augmented-MBPP -- indicating an over 60\% improvement in distributional alignment for both benchmarks.} 

\sloppy \smallskip \noindent \textit{By improving their KU coverage and distributional alignment, all studied LLMs exhibit a consistent performance drop on the augmented benchmarks that is statistically significant with a large effect size. The performance drop ranges from 12.54\% to 44.82\% across all evaluation levels (e.g., pass@1, pass@3 and pass@5), highlighting that under-evaluated KUs pose significant challenges for LLMs, likely because current benchmarks give limited attention to these KUs. We also find that each model exhibits a unique profile of KU-specific strengths and weaknesses, reinforcing the importance of KU-level evaluation in uncovering nuanced model capabilities that current benchmarks (e.g., HumanEval and MBPP) may not fully capture.}



\medskip \noindent \textbf{The main contributions of our paper are as follows:} 
(i) we present the first study to analyze the coverage of language concepts of programming languages in benchmark datasets through the lens of KUs and identify major gaps compared to KU usage in real-world software projects,
(ii) we design and prototype a framework to automatically generate KU-based tasks that can improve the representativeness of KUs in existing benchmarks, and
(iii) we conduct a thorough empirical evaluation of seven popular LLMs, understanding their strengths and weaknesses across different KUs. 

\smallskip
\noindent Our framework and findings provide actionable insights for both researchers and practitioners. They can be used to (a) assess and diagnose language concept coverage in existing benchmarks, (b) construct more comprehensive and realistic benchmark suites, and (c) guide the development of future LLMs that are evaluated—and ultimately improved—across the full conceptual space of programming-language concepts. A supplementary material package is provided online\footnote{\url{https://spplementary_package_url}}.

\smallskip \noindent \textbf{Paper organization.} Section~\ref{sec:Knowledge_Unit} defines KUs and presents our approach for detecting them. Section~\ref{sec:data_collection} describes our data collection and preparation approach. Section~\ref{sec:Main_study} presents the motivation, approach, and findings of our research questions. Section~\ref{sec:Discussion} discusses the implications of our study. Section~\ref{sec:Related_Work} discusses related work. Section~\ref {sec:Limitations_And_Threats} describes the threats to the validity of our findings. Finally, Section~\ref{sec:Conclusion} outlines our concluding remarks.

	\section{Knowledge Units (KUs)}
\label{sec:Knowledge_Unit}

In our prior work, we proposed knowledge units (KUs) of programming languages to analyze the source code of software systems for a variety of software analytics tasks~\citep{ahasanuzzaman2024using, ahasanuzzaman2025predictingLTC,ahasanuzzaman2024predictingDefect}. As we will use this notion in this paper as a lens to model language concepts, Section~\ref{subsec:knowledge_definition} provides a quick overview of KUs to make this paper as self-contained as possible. Additionally, we describe how we leverage  large language models (LLMs) to elicit KUs in Section~\ref{subsec:knowledge_operational_definition} and detect them in Section~\ref{subsec:knowledge_detection}.


\subsection{Definition}
\label{subsec:knowledge_definition}

We define a Knowledge Unit (KU) as a cohesive set of key capabilities that are offered by one or more building blocks of a given programming language. For the Python programming language, these building blocks encompass both the Python fundamental language constructs (e.g., if/elif/else statements, async/await  statements and with statement) and APIs (e.g., Concurrency API, IO API, Regular Expression API and String API). Each building block offers a \textit{set of capabilities} essentially representing the actions a developer can perform using that building block. The inclusion of ``key'' in our definition aims to ensure that KUs are centered around fundamental capabilities rather than those that are overly specific.  

\begin{figure}[!t]
    \centering
    \includegraphics[width=1.0\textwidth]{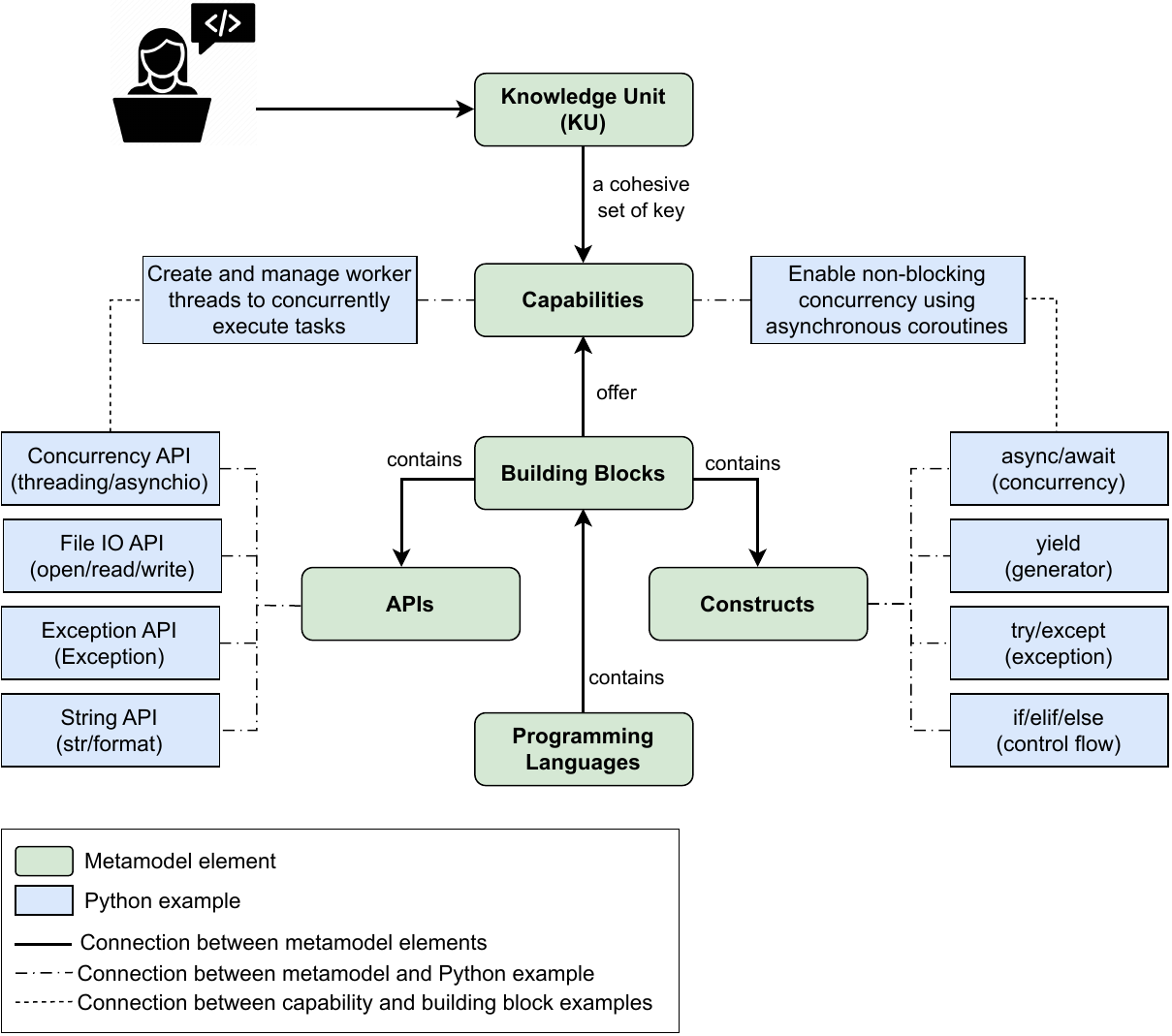}
    \caption{Our metamodel for knowledge units (KUs).}
    \label{fig:ku_metamodel}
\end{figure}

Figure~\ref{fig:ku_metamodel} illustrates a subset of key capabilities that are associated with the Concurrency APIs (left-hand side) and language constructs (right-hand side). Using the Python constructs such as \texttt{async} or \texttt{await} developers can enable non-blocking concurrency using asynchronous coroutines, whereas with Concurrency APIs (e.g., \texttt{threading} or \texttt{asyncio}) developers can create and manage worker threads to concurrently execute tasks. Thus, the Concurrency KU encapsulates a cohesive set of key concurrency-related capabilities that are offered by the programming language's underlying building blocks (i.e., APIs and constructs).

\subsection{Operationalizing KUs via LLMs}
\label{subsec:knowledge_operational_definition}
To objectively quantify KUs in a programming language, an operational definition for KUs is needed. In our previous works~\cite{ahasanuzzaman2024using,ahasanuzzaman2025predictingLTC,ahasanuzzaman2024predictingDefect}, we used Oracle Java certification exams (such as Oracle Java SE and Java EE exams) to operationalize KUs for the Java programming language. However, in this paper, we leverage large language models (LLMs) to operationalize KUs for the Python programming language. Our rationale for choosing LLMs is two-fold: (1) they can provide richer and more diverse information compared to a single source, such as certification exam websites. LLMs are trained on extensive datasets that include web pages, books, source code, and documentation, offering a broader perspective on programming language concepts~\citep{minaee2024large}; and (2) they support multiple programming languages, whereas prior works~\cite{ahasanuzzaman2024using,ahasanuzzaman2025predictingLTC,ahasanuzzaman2024predictingDefect} were limited to a single language (e.g., Java). This multi-language capability enables a more comprehensive and generalizable analysis of programming knowledge.

\noindent\textbf{Approach:} We briefly explain each step that we follow to operationalize KUs using LLMs.
\begin{itemize}[wide = 0pt, label=$\bullet$]
    \item \textit{Step 1) Generate a prompt template for operationalizing KUs:} We generate a one-shot prompt template to ask the LLM to list all KUs of the Python programming language along with their key cohesive capabilities. Figure~\ref{fig:ku_oper_prompt} in Appendix~\ref{appendix:prompt_tamplate} illustrates our one-shot template, which consists of two main parts. In the `Context' section, we provide the general concept of a KU to guide the model's output effectively. This section also includes an example that demonstrates the Loop KU of Python language and its associated key capabilities. We craft this example based on the list of KUs and their capabilities presented in our previous work~\cite{ahasanuzzaman2024using}. This example helps to direct the model's output. In the `Instruction' section, we ask the model to list all Python programming language KUs and their key cohesive capabilities. We also include a presentation guideline that instructs the model to follow the format provided in the example.

    \item \textit{Step 2) Execute the prompt template using the selected LLM:}  
    We adopt a dialogue-based approach to iteratively refine and expand the generated outputs. Specifically, we select \texttt{ChatGPT} as our LLM for detecting KUs, as it has demonstrated superior performance in software engineering tasks~\citep{DBLP_conf_acl_GaoFC20, min_etal_2022_rethinking, white2023prompt, zhou2022large}. We use the \texttt{GPT--4o--Mini} variant, which at the time of writing is the most cost-efficient model that surpasses \texttt{GPT--3.5--Turbo} and other smaller models in both text-based and multimodal academic benchmarks~\citep{chat_gpt_4o_mini_adventage}. The model version \texttt{gpt-4o-mini-2024-07-18} is used with a temperature setting of 0.5 to maintain a balance between creativity and determinism in output generation~\citep{jain2023improving,zhang2025thinking}.  
    
    After executing the initial prompt, we continue the process using a dialogue-based refinement strategy. We prompt the model with a follow-up question: ``Can you find more KUs of the programming language and their associated key capabilities?'' This dialogue continues iteratively to identify additional KUs that may have been overlooked in previous responses. We repeat this dialogue until we observe that the generated list begins to include KUs or capabilities already mentioned in previous iterations. In our approach, we repeat this dialogue up to five times, after which we consistently observe repetitions of previously identified KUs or capabilities and terminate this process.
    
    \item \textit{Step 3) Analyze the outputs of the selected LLM:} The first and third authors jointly review and analyze the outputs of the model. They specifically examine whether the generated KUs and capabilities are related to Python programming language. The authors discuss together and resolve any issues. During this analysis, the authors observe that the model sometimes outputs KUs related to third-party frameworks. For instance, the model generates concepts related to \texttt{pandas} (with capabilities describing how to manipulate data using the pandas library) and KUs related to \texttt{scikit-learn} (with capabilities focused on using scikit-learn APIs for model training, testing, and evaluation). 
    \begin{table}[!htbp]
    \centering
    \caption{The identified Python programming language knowledge units (KUs).}
    \label{tab:topic_definition}
    \resizebox{\columnwidth}{!}{
    \begin{tabular}{p{3.2cm}p{9cm}}
        \toprule
        \multicolumn{1}{C{3cm}}{\textbf{Knowledge unit (KU)}} & \multicolumn{1}{C{9cm}}{\textbf{Definition}}                                                                                                                                                                                                                               \\ \midrule

        \textbf{[K1]} Variable                                         &
        The ability to declare and assign variables using Python's built-in data types (e.g.,  integers, floating-point numbers, strings and the special None type)
        \\ \midrule

        \textbf{[K2]} Operators                             &
        The use of Python operators to perform arithmetic, comparison, logical, assignment, bitwise and membership operations to manipulate and evaluate data.
        \\ \midrule

        \textbf{[K3]} Condition                                            &
        The ability to control program flow using conditional statements including nested conditionals, short-circuit evaluation, and Python's ternary operator.
        \\ \midrule

        \textbf{[K4]} Loop                                             &
        The ability to execute repetitive tasks using for and while loops, use loop control keywords like pass and continue, and apply while-else and for-else clauses.
        \\ \midrule

        \textbf{[K5]} Function                         &
        The definition and invocation of reusable code blocks using \texttt{def}, passing arguments (positional, keyword, default), returning values, and managing variable scope (local vs. global)
        \\ \midrule

        \textbf{[K6]} Anonymous\newline Function                                      &
        The use of lambda expressions to create small, unnamed functions for short, inline operations without formally defining a function using \texttt{def}.
        \\ \midrule

        \textbf{[K7]} Data Structure                             &
        The creation and manipulation of Python’s built-in collection types — lists, dictionaries, sets, and tuples — including indexing, slicing, appending, updating, and iterating over elements.
        \\ \midrule

        \textbf{[K8]} File Handling                           &
        The ability to open, read, write, and close files using Python's built-in \texttt{open()} function and file object methods like \texttt{read()}, \texttt{readline()}, \texttt{write()}, and \texttt{writelines()}
        \\ \midrule

        \textbf{[K9]} Object-Oriented Programming                         &
        The use of Python's class-based structure to define and instantiate objects, encapsulate data and behavior, use different types of methods (instance, class, static), apply inheritance for reuse, and support polymorphism.
        \\ \midrule

        \textbf{[K10]} Exception\newline Handling                                      &
        The ability to handle runtime errors gracefully, raise exceptions, define custom exception classes, handle multiple exceptions, and cleanup tasks after exception is handled.
        \\ \midrule

        \textbf{[K11]} Generators                                      &
        The use of generator functions, defined with the \texttt{yield} statement, to create iterators that lazily produce a sequence of values, optimizing memory usage and enabling iteration with functions like \texttt{next()} and \texttt{iter()}.
        \\ \midrule

        \textbf{[K12]} Decorators                                   & 
        The programming construct used to modify or extend the behavior of functions or methods without altering their source code. This includes the use of built-in decorators such as \texttt{@staticmethod}, \texttt{@classmethod}, and \texttt{@property}, as well as the creation and application of custom decorators, with or without parameters.
        \\ \midrule

        \textbf{[K13]} Closures                                              &
        The creation of functions that capture and remember values from their enclosing scope by defining a nested function inside another function and returning it, forming a closure
        \\ \midrule

        \textbf{[K14]} Context\newline Managers                                              &
        The use of the with statement to manage resources (e.g.,  files or network connections), ensuring they are properly acquired and released, either through built-in support or custom context manager classes using contextlib module.
        \\ \midrule

        \textbf{[K15]} Comprehension                                &
        The concise creation of data structures using list, dictionary, and set comprehensions, as well as generator expressions, often used for transforming or filtering data in a single line.
        \\ \midrule

        \textbf{[K16]} Concurrency                                      &
        Writing programs that can execute multiple tasks simultaneously using Python's threading, multiprocessing, async/await syntax, and concurrency modules such as \texttt{concurrent.futures}.
        \\ \midrule

        \textbf{[K17]} String\newline Manipulation                                        &
        The ability to create, modify, and analyze strings using string methods (\texttt{upper()}, \texttt{lower()}, etc.), slicing, indexing, string formatting (including f-strings), and apply regular expression and pattern matching for parsing or searching.
        \\ \midrule   

        \textbf{[K18]} Networking                                     &
        The ability to implement network communication using low-level sockets or high-level libraries like requests and urllib, including client-server interactions and HTTP protocol usage.
        \\ \midrule

        \textbf{[K19]} Serialization                                  &
        The ability to convert data between in-memory Python objects and serialized formats such as JSON, XML, or binary (using pickle) for storage or transmission.                                                             \\ \midrule

        \textbf{[K20]} Database                            &
        The ability to perform database operations using \texttt{sqlite3} or Python Object-Relational Mappers (ORMs) like \texttt{SQLAlchemy}, including connecting to databases, running queries, and managing transactions.                                                                                                                                                                                                              \\  \bottomrule
    \end{tabular}
    }
\end{table}
    As in this paper, our goal is to identify general-purpose Python KUs that reflect the language’s built-in features and conceptual constructs such as concurrency, file handling, or exception management -- rather than usage patterns tied to third-party frameworks, the authors decide to remove such KUs that are specific to external frameworks or libraries.
    
    Finally, the LLM-based approach ends up with 20 Python programming language KUs and their associated capabilities from the model's output. Table~\ref{tab:topic_definition} lists the identified 20 Python KUs and their definition. The list of key capabilities that are associated with each of the identified KUs is provided in Table~\ref{tab:ku-capabilities} in Appendix~\ref{appendix:python_kus}.
    
    We further validate the completeness of the identified KUs by cross-checking them against the topics covered in the Python Certification exams offered by the \textit{Python Institute}\footnote{\url{https://pythoninstitute.org/about-pi}} The Python Institute offers four certification levels: Entry-Level (PCEP)~\citep{pyinst_pcep}, Associate-Level (PCAP)~\citep{pyinst_pcap}, Professional-Level 1 (PCPP1)~\citep{pyinst_pcpp1}, and Professional-Level 2 (PCPP2)~\citep{pyinst_pcpp2}. The authors carefully examine all topics covered across these exams and observed that the core and advanced concepts evaluated in these certifications are fully represented within the generated KUs and their corresponding capabilities. Interestingly, we also observe that LLMs generate a few additional language concepts such as generators, closures, and decorators that are not explicitly included in the certification exam topics. This observation further supports the comprehensiveness of our KU set, as it not only captures the core concepts taught and tested in formal certifications but also includes practical and advanced features frequently encountered in real-world Python programs (e.g., closures are crucial for callback functions in event-driven programming~\citep{Benfield2019ExpertT}).

\end{itemize}

\subsection{Detection of the incidence of KUs via LLMs}
\label{subsec:knowledge_detection}

To detect the incidence of KUs in a Python file, we again leverage LLMs. Our rationale for choosing LLMs over static analysis tools is two-fold: static analysis tools are costly to develop and language-specific, and they often struggle to capture the comprehensive code context. In contrast, LLMs are more cost-effective to build and offer a more nuanced insights and suggestions by considering the broader code context. 

\noindent\textbf{Approach:} We use a prompt engineering technique to detect the incidence of Python KUs from a given piece of code (e.g., .py file). First, we generate a prompt template to instruct an LLM for detecting KUs. Next, we execute the prompt template using \texttt{Open AI} APIs for the (\texttt{GPT--4o--Mini}) model. Finally, we collect the count of the incidence of KUs that are present in the file using our custom parser. Below, we briefly describe every step that we follow to detect the incidence of KUs.

\begin{itemize}[wide = 0pt, label=$\bullet$]
    \item \textit{Step 1) Generate a prompt template for detecting the incidence of KUs:} To detect the incidence of a KU, we generate a clear and concise prompt template (see Figure~\ref{fig:ku_detection_prompt} in Appendix~\ref{appendix:prompt_tamplate}). In this prompt template, we include context about KUs, provide instruction for the LLM, add example inputs and outputs,  and insert the query code. In the following section, we briefly explain each step to generate our prompt template.
        \sloppy
        \begin{itemize}[wide = 0pt, itemsep = 3pt, topsep=3pt, listparindent=\parindent]
            \item \textit{Step 1.1) Include context about KUs}. We include context about KUs at the beginning of our prompt template (see the `\#Context' section in Figure~\ref{fig:ku_detection_prompt}). This context includes the general concept of a KU, the list of identified 20 KUs and their capabilities. First, we clearly mention that the LLM is expert at analyzing program. Next, we include the definition of our KUs that helps the LLM understand the concept of KUs. Finally, we state that the LLM is familiar with 20 KUs of the Python programming language which are delimited by `[py-kus]' tags. Within these tags, we list the names of the KUs and their capabilities. Each capability of a KU is separated by a comma. Including this list of KUs and their capabilities ensures that the provided prompt allows us to retrieve all KUs present in a given piece of code. It also constrains the model’s responses, ensuring that its outputs remain relevant and aligned with the provided information.
            
            \item \textit{Step 1.2) Provide instruction for the LLM.} In the `\#Instruction' section of the prompt template, we direct the LLM to detect the instances of all KUs from a given piece of code, focusing on the KUs listed within the `[py-kus]' tags (see Step 1.1). In particular, we ask the model to count the occurrences for each KU capability present in the given piece of code. The model also counts each capability of a KU. This count is the total number of instances of a KU present in the given piece of code. 
            
            To illustrate how this process works, consider the Concurrency KU, which consists of four key capabilities (see Table~\ref{tab:ku-capabilities}): [C1] managing concurrent tasks using the threading module to run multiple threads within a single process, [C2] achieving parallel execution with the multiprocessing module by creating separate processes, [C3] writing asynchronous code using the async and await keywords for cooperative multitasking and [C4] using the concurrent.futures module to simplify concurrent task execution. Suppose the given piece of code creates one \texttt{thread} instance [C1], defines two asynchronous functions using \texttt{async} [C3], and executes one task using the \texttt{concurrent.futures} module [C4]. In this case, the model detects 1 occurrence of [C1], 2 occurrences of [C3], and 1 occurrence of [C4]. Thus, the total number of occurrences for the Concurrency KU in this code is therefore 1 + 2 + 1 = 4. 
            
            \item \textit{Step 1.3) Add example inputs and outputs.} To guide the model in understanding how to detect KUs and how to format the output appropriately, we manually generate two code examples and their output (i.e., the count of the instances of all KUs present in the code) in JSON format. We include these two examples under the \textit{``\# Few shot examples''} section of the prompt template. Each example consists of an input code snippet placed below the \textit{``Input:''} label, followed by the corresponding output shown below the \textit{``Output:''} label. The output contains the list of all KUs along with their respective total counts of their instances in JSON format.
            
            \item \textit{Step 1.4) Insert the query code.} We include the query code using the delimiter ``query\_code'' -- the specific piece of code for which the LLM needs to detect the incidence of KUs. We place a question mark (`?') next to the \textit{``Output:''} label to indicate that we request the LLM to generate the output, which includes the names and counts of the instances of the identified KUs from the query code.
        \end{itemize}
        
    \item \textit{Step 2) Execute the prompt template:}  We use the \texttt{gpt-4o-mini-2024-07-18} variant of the \texttt{GPT--4o--Mini} model via the \texttt{OpenAI} Batch APIs to execute the prompt template. Our rationale for using the Batch APIs is that they allow us to send multiple requests asynchronously, reducing costs by 50\%.  We create batches of 200 prompts, send them via Batch APIs to ChatGPT, and store the responses as JSON objects. In this context, the temperature parameter is set to 0.2, which reduces randomness and promotes consistent and deterministic results~\citep{open_ai_temperature}.
    
    \item \textit{Step 3) Collect the  list of detected KUs along with their capabilities:}  We parse the JSON file for every response. We extract the name of the identified KUs and the count of the incidence of every identified KU capability from each response of the LLM. 

    To validate the performance of our LLM-based KU detector, the first and third authors manually label KUs in the solution code of the HumanEval datasets and 200 randomly selected files from the studied real-world Python projects. The authors work together during the labeling process and discuss any ambiguous cases to reach an agreement. We use these manually labeled datasets as ground truth and evaluate the accuracy of the KU detector. We observe that our LLM-based KU detector achieves an accuracy of 93\% in detecting KUs.
\end{itemize}

    \section{Data Collection and Preparation of the Empirical Study}
\label{sec:data_collection}

In this section, we describe the dataset collection and preparation process used to address the research questions outlined in the introduction. Figure~\ref{fig:data_collection} provides an overview of this process. First, we collect the benchmark datasets (Step 1). We then extract instances of KUs from each task in benchmark datasets (Step 2). To enable a comparison between benchmark tasks and real-world projects through the lens of KUs, we then download the source code of the studied projects (Step 3). Finally, we extract instances of KUs from the source code of the studied projects (Step 4). At the end of this process, we obtain KU datasets for both benchmarks and real-world projects that we use in our study. In the following, we describe each step of our data collection and preparation process in detail.



\begin{itemize}[wide = 0pt, label=$\bullet$]
    \item \textit{Step 1) Collect benchmark datasets for code generation tasks:} We choose two benchmark datasets (1) HumanEval~\citep{chen2021evaluating} and (2) MBPP~\citep{austin2021program} because they are the most widely used benchmark datasets~\citep{zhang2025thinking, jiang2024survey, shinn2023reflexion, islam2025codesim, wang2023recode, Nijkamp2023CODEGENAL} for evaluating the performance of LLMs in code generation tasks. We downloaded these datasets from their respective official GitHub repositories for HumanEval~\footnote{\url{https://github.com/openai/human-eval}} and MBPP~\footnote{\url{https://github.com/google-research/google-research/tree/master/mbpp}}.

    The HumanEval dataset consists of 164 handwritten Python programming tasks. Each task includes a function signature, docstring, body, and several unit tests. The MBPP dataset consists of 974 short Python problem tasks created through crowdsourcing from an internal pool of crowdworkers with knowledge of Python. The natural language descriptions for each task are typically short, usually one sentence. Both the HumanEval and MBPP datasets include canonical solutions for each task, which are written by humans. We use these canonical solutions to understand what programming language KUs are needed to solve each task in code generation benchmarks.

\begin{figure}[!t]
    \centering
    \includegraphics[width=1.0\textwidth]{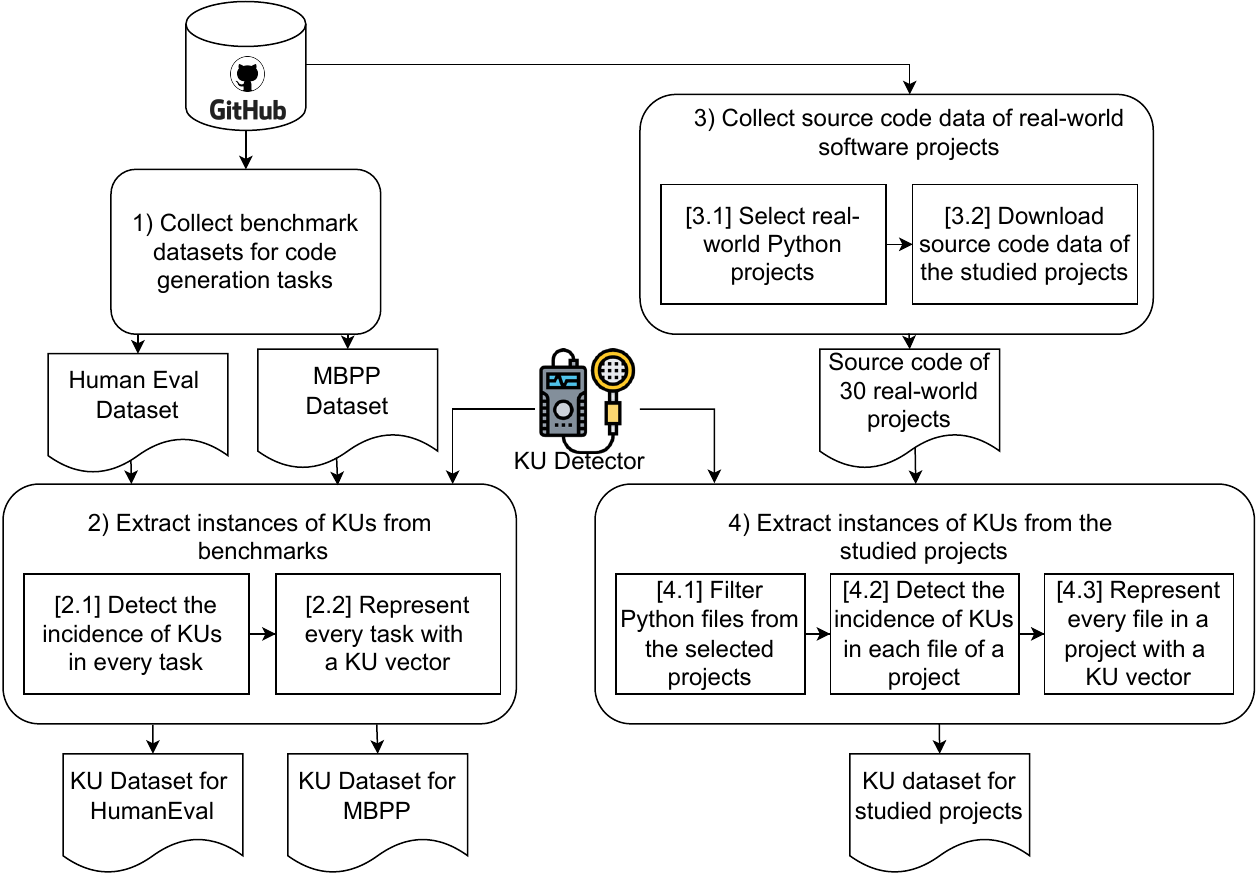}
    \caption{An overview of our data collection and preparation process.}
    \label{fig:data_collection}
\end{figure}

    \item \textit{Step 2) Extract instances of KUs from benchmarks:}
        \begin{itemize}[wide = 0pt, itemsep = 3pt, topsep=3pt, listparindent=\parindent]
            \item \textit{Step 2.1) Detect the incidence of KUs in every task}. Both the HumanEval and MBPP datasets include canonical solutions for each task, which are written by humans. We use these canonical solutions to understand what programming language KUs are needed to solve a code generation task. We detect the incidence of KUs from each task of a benchmark using the LLM-based KU detector (c.f., Section~\ref{subsec:knowledge_detection}). 
            For each task, the canonical solution code is used as the query input within the KU detection prompt template. The model then analyzes this input and produces a structured JSON response, listing the detected KUs and their corresponding capabilities. Finally, we parse this JSON output to count the number of occurrences (incidences) of each KU capability identified in the task. As soon as one capability of a given KU is found, we consider this KU to be covered by the analyzed source code.
            
            \item \textit{Step 2.2) Represent every task with a KU vector}. For every task in a benchmark, we first sum the total number of occurrences (incidences) of all capabilities belonging to each KU. We then represent the task as a 20-dimensional KU vector, where each dimension corresponds to a specific KU and stores the sum value of its occurrences within that task. Finally, for each benchmark, we build a dataset consisting of these KU vectors, with each vector representing the KU distribution of an individual task.
        \end{itemize}
    \item \textit{Step 3) Collect source code data of real-world software projects:} 
    \begin{itemize}[wide = 0pt, itemsep = 3pt, topsep=3pt, listparindent=\parindent]
        \item \textit{Step 3.1) Select real-world Python projects}. Our selected benchmark datasets contain code generation tasks for the Python programming language. Therefore, we focus our study on real-world Python projects. Specifically, we consider popular and actively maintained Python projects that are hosted on GitHub. However, there exist toy projects, personal projects, academic projects, and websites on GitHub that fulfill those criteria~\citep{promise_perils_GitHub_MSR_2014}. Hence, in order to select real-world software projects, we leverage the curated list of engineered projects that are \textbf{manually verified} by Muniay et al.~\citep{engineered_project_github_EMSE_2017}. This dataset, commonly known as RepoReapers, has been widely adopted in recent software engineering studies for selecting subject systems~\citep{sharma2023investigating, alghamdi2021characterising, wessel2023github, synovic2022snapshot, babur2024language, sharma2024multi}. This dataset contains both organizational and utility projects that represent real-world software projects. 
        
        Since our focus is on Python KUs, we include only those projects for which Python is the primary programming language.  From this filtered set, we identify 46 Python projects, of which 15 are organizational projects and 31 are utility projects. To maintain a balanced and unbiased analysis across different project types, we select 15 organizational and randomly select 15 utility projects, resulting in a total of 30 Python projects used for our study.
    
        \item \textit{Step 3.2) Download source code data of the studied projects}. To download the current snapshot of each software project at the time of data collection (November 10th,  2024), we cloned the GitHub repository of each project using the \texttt{git clone} command. The complete list of selected projects along with their corresponding GitHub repository links is provided in our supplementary materials.
    \end{itemize}
    
    \item \textit{Step 4) Extract instances of KUs from the studied projects:}
        \begin{itemize}[wide = 0pt, itemsep = 3pt, topsep=3pt, listparindent=\parindent]
            \item \textit{Step 4.1) Filter Python files from the selected projects.} To extract instances of KUs from the Python source code, we select files with the \texttt{.py} extension. We exclude \texttt{\_\_init\_\_.py} files as they are primarily used for package configuration rather than containing executable code. We did not filter out test files, since we found that developers also exercise KUs when writing and maintaining test suites and initial analysis of the data after filtering out test files did not show changes in the proportion of detected KUs. Finally, we select 7,573 Python files from the selected projects.
            \item \textit{Step 4.2) Detect the incidence of KUs in each file of a project.} We detect the incidence of Knowledge Units (KUs) in each source file of a project using the LLM-based KU detector (see Section~\ref{subsec:knowledge_detection}). Specifically, the content of each source file is used as the query input within the KU detection prompt template. To ensure that every file fits within the model’s token limit (e.g., 128K for GPT-4o-mini), we preprocess each file by removing comments and docstrings before sending it to the LLM. 
            The model then analyzes the file and returns a structured JSON response containing the identified KUs and their associated capabilities. Finally, we parse this JSON output to extract and count the number of incidence for each detected KU capability present in the file.
            
            \item \textit{Step 4.3) Represent every file in a project with a KU vector.} For every source file, we sum the total number of occurrences (incidence) of all capabilities belonging to each identified KU. We then represent every file as a 20-dimensional KU vector, where each dimension corresponds to a specific KU and stores the total count of its occurrences within that file. Finally, for each project, we build a dataset composed of these KU vectors, with each vector representing the KU distribution of an individual source file.
        \end{itemize}
\end{itemize}

Having constructed the KU datasets for both benchmarks and real-world Python projects, we now have a unified representation that enables a direct comparison across benchmarks and projects through the lens of KUs. Building on these datasets, the next section presents our methodology for addressing the research questions (RQs) in this empirical study.

    \section{Empirical study of the language concepts used in code generation}
\label{sec:Main_study}

In the following, we address our two research questions. For each research question, we discuss our motivation for studying it, the approach used to answer it, and the findings that we observe.

\subsection{RQ1: \RQOne}
\label{sec:rq1}

To assess how comprehensively existing LLM benchmark datasets cover language concepts compared to real-world software projects, we investigate this research question from two complementary perspectives. First, we analyze KU coverage by measuring the proportion of instances of each KU that appear in benchmark datasets relative to their presence in real-world projects. Second, we analyze the distributional balance of KUs to determine whether benchmarks emphasize certain KUs disproportionately. Together, these two perspectives provide a comprehensive view of the conceptual completeness and representativeness of benchmarks in terms of KUs. In particular, we study the following two sub-research questions:

\begin{itemize}
  \item \textbf{RQ1.1:} \RQOneSubOne
  \item \textbf{RQ1.2:} \RQOneSubTwo
\end{itemize}

\subsubsection{RQ1.1: \RQOneSubOne}
\label{sec:RQ1_1}
\vspace{0.2cm}

\noindent{\textbf{Motivation:}} Benchmarks such as HumanEval and MBPP are used to evaluate the performance of large language models (LLMs) on code generation tasks. However, it remains unclear whether these benchmarks adequately reflect the diversity of programming language KUs encountered in real-world software development. By comparing KU coverage in benchmark datasets with that in real-world projects, we aim to identify potential gaps in the current benchmark design.

\vspace{0.2cm}
\noindent{\textbf{Approach:}}
To investigate RQ1.1, we measure the KU coverage in both benchmark datasets and real-world projects, and compare them to evaluate how well benchmarks reflect real-world KU distributions.

\noindent\textbf{Quantify KU coverage.}
To quantify the coverage of KUs in a benchmark or project, we calculate the percentage of instances of each KU for benchmark data sets and real-world projects. We compute the coverage of a KU \textit{K} using the following formula:

\[
\textstyle
\text{Coverage of a KU } K = \left( \frac{\text{Total instances of KU } K \text{ in the dataset}}{\text{Total instances of all KUs in the dataset}} \right) \times 100\%
\]
Here, the dataset refers to either a benchmark (e.g., HumanEval or MBPP) or a real-world software project.

\noindent\textbf{Compare KU coverage between two categories of real-world projects.} Our dataset includes two categories of real-world projects --  organizational and utility. To statistically determine whether the coverage of any KU differs between these categories, we apply the Wilcoxon rank-sum test (also known as the Mann--Whitney U test)~\citep{whitney_test}, a non-parametric statistical method for comparing two independent groups. Since we perform the test independently for each KU, this results in multiple comparisons (e.g., 20 Mann–Whitney comparison tests between the two project categories for 20 KUs), increasing the risk of false positives (Type I error). To control for multiple comparisons, we apply Bonferroni correction~\citep{bonferroniTest}, which adjusts the significance threshold to $\alpha$= 0.0025 (0.05/20) to reduce the likelihood of false positives (Type I errors).

\noindent\textbf{Compare KU coverage between real-world projects and a benchmark.} To statistically assess whether KU coverage differs between real-world projects and a benchmark dataset (e.g., HumanEval or MBPP), we apply the Wilcoxon signed-rank test -- a non-parametric method for comparing two dependent groups. Specifically, we compute the median KU coverage for each KU across the real-world projects to form a distribution. We then perform a paired Wilcoxon signed-rank test  between the distribution of KU coverage in real-world projects and that of the benchmark dataset, with each pair formed by the same KU across the two groups. Since we perform this comparison for multiple benchmarks, we apply Bonferroni correction to adjust the significance threshold to $\alpha$= 0.025 (0.05/2) and control for potential false positives due to multiple testing.

\vspace{0.2cm}
\noindent{\textbf{Findings:}}
\begin{figure}[!t]
	\centering
	\includegraphics[width=1\linewidth]{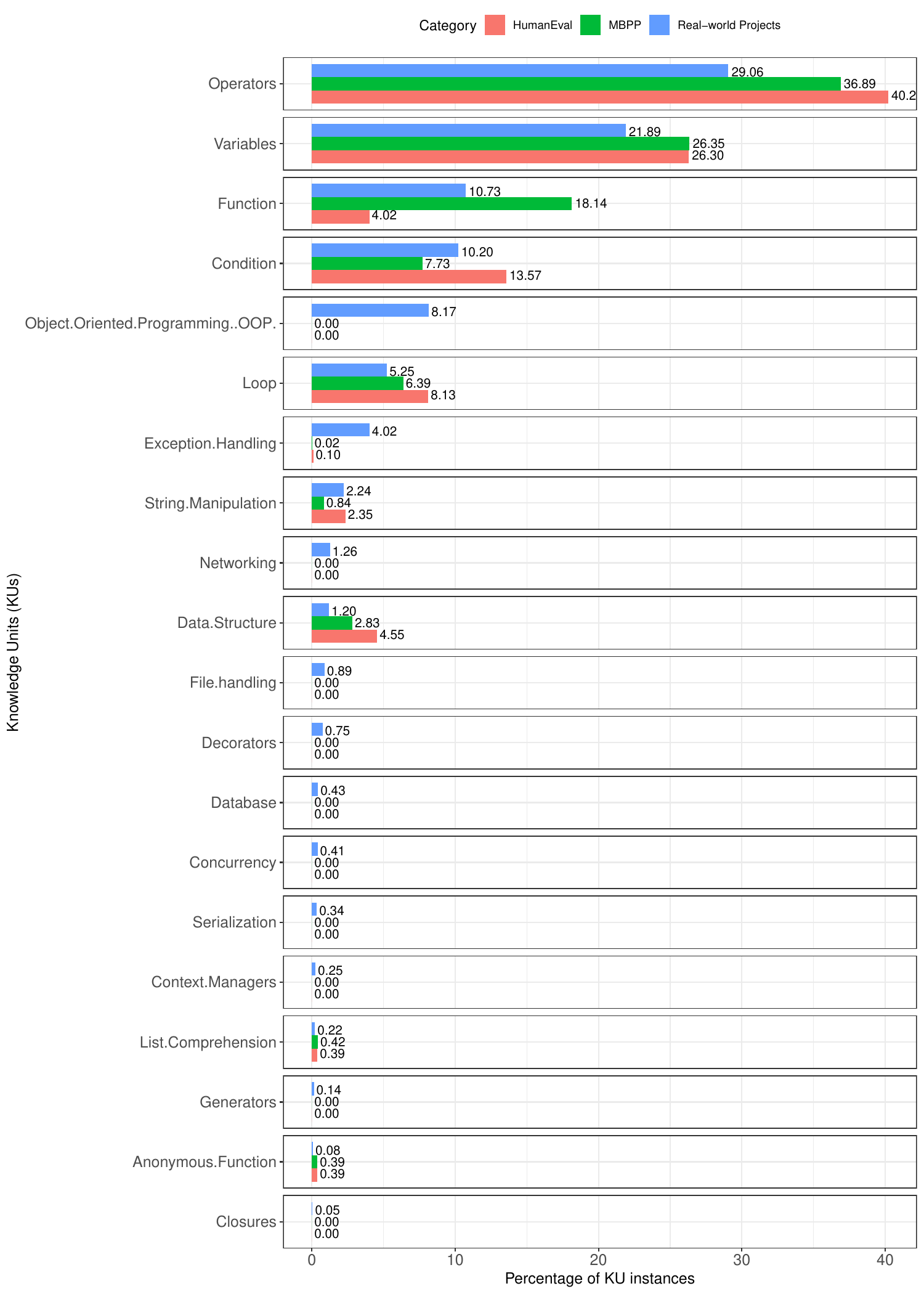}
	\caption{Percentage of instances of every KU across HumanEval, MBPP and real-world projects.} 
	\label{fig:rq1_ku_ratio}
\end{figure}
\smallskip \observation{Only 50\% (10 out of 20) of the identified KUs are present in each of the HumanEval and MBPP datasets and both benchmarks include the same set of KUs, indicating that neither benchmark provides comprehensive coverage of the key KUs of the Python programming language.} Figure~\ref{fig:rq1_ku_ratio} presents the percentage of instances of every KU across HumanEval and MBPP. In both benchmark datasets, only half of the identified KUs are present (i.e., the percentage of instances for 50\% of the identified KUs is greater than zero). Moreover, the set of identified KUs is identical in both benchmarks. We observe that Operators and Variables are the most frequently observed KUs in both datasets, whereas many KUs such as List Comprehension, Anonymous Function, and Exception Handling KUs appear infrequently in both datasets. For example, the KU coverage for Operators is 40.2\% for HumanEval and 36.9\% for MBPP. On the contrary, Exception Handling KU accounts for only 0.1\% of KU instances in HumanEval and 0.02\% in MBPP. We also observe notable differences between these benchmark datasets in terms of KU coverage. The Function KU appears more frequently in MBPP, whereas Loop, String Manipulation, Data Structure, and Condition KUs are more prevalent in HumanEval. Importantly, 50\% of the identified KUs are absent from both benchmark datasets. Many of these missing KUs are fundamental to Python programming, including Object-Oriented Programming, Generators, and Context Managers KUs. The remaining seven missing KUs are as follows: File Handling, Closures, Decorators, Concurrency, Networking, Database, and Serialization KUs.

\smallskip \observation{Real-world software projects make use of all 20 identified KUs of the Python programming language.} Figure~\ref{fig:rq1_ku_ratio} also presents the median percentage of instances of every KU in real-world projects. We observe that all identified KUs are present in real-world projects (i.e., the KU coverage for every KU is greater than zero). Notably, every KU appears in both organizational and utility project categories. We observe no  statistically significant differences of each KU coverage between organizational and utility projects, indicating that KU distributions are consistent in both project categories.


Among the identified KUs present in real-world projects, Operators and Variables are the most prevalent Python KUs (they together account for almost 50\% of all KU instances). The fundamental Python KUs such as Object-Oriented Programming and Exception Handling are also well represented, contributing 8.17\% and 4.02\% of the KU instances, respectively. While File Handling, Serialization, Database, and Concurrency KUs each constitute less than 1\% of the overall KU instances, they reflect advanced and critical programming capabilities in Python. We also find that Anonymous Function, Closures, and Generators appear infrequently, yet they represent essential KUs of the Python. These results underscore that real-world projects exercise the full spectrum of Python KUs.


\smallskip \observation{Benchmark datasets do not follow the same KU distribution trends as real-world projects.} Table~\ref{tab:comp_ku_proj_bench} presents the comparison of the KU coverage between real-world projects and benchmark datasets. The last two columns show the absolute difference in the KU coverage, where a positive value indicates a higher percentage of KU instances in the benchmark than in real-world projects, and a negative value indicates a lower percentage of KU instances.

\begin{table}[]
\centering
\rotatebox{90}{
\begin{minipage}{\textheight}
\caption{Comparison of KU coverage across real-world projects, HumanEval, and MBPP datasets. A Negative difference indicates that a benchmark has a lower KU coverage ($\downarrow$) than real-world projects, whereas a positive difference indicates that the benchmark has higher KU coverage ($\uparrow$) than real-world projects.}
\label{tab:comp_ku_proj_bench}
\resizebox{\columnwidth}{!}{
\begin{tabular}{@{}lrrrrrrr @{}}
\toprule
\multicolumn{1}{c}{}                                                                                            & \multicolumn{3}{c}{\textbf{\begin{tabular}[c]{@{}c@{}}KU coverage (\%) (median) \\ for software projects\end{tabular}}}                                                                                                                                                                                                 & \multicolumn{2}{c}{\textbf{\begin{tabular}[c]{@{}c@{}}KU coverage (\%)\\ for benchmarks\end{tabular}}} & \multicolumn{2}{c}{\cellcolor[HTML]{FFFFFF}\textbf{\begin{tabular}[c]{@{}c@{}}Difference of KU coverage (\%)\\ between Benchmarks and \\ Real-World  Projects\end{tabular}}}                                                                         \\ \cmidrule(l){2-8} 
\multicolumn{1}{c}{\multirow{-3}{*}{\textbf{\begin{tabular}[c]{@{}c@{}}Knowledge Units \\ (KUs)\end{tabular}}}} & \multicolumn{1}{c}{\textbf{\begin{tabular}[c]{@{}c@{}}Organizational\\ Projects\end{tabular}}} & \multicolumn{1}{c}{\textbf{\begin{tabular}[c]{@{}c@{}}Utility\\ Projects\end{tabular}}} & \multicolumn{1}{c}{\textbf{\begin{tabular}[c]{@{}c@{}}Real-World\\ Projects \\ (Organizational \\ \& Utility)\end{tabular}}} & \multicolumn{1}{c}{\textbf{HumanEval}}                    & \multicolumn{1}{c}{\textbf{MBPP}}                    & \multicolumn{1}{c}{\textbf{\begin{tabular}[c]{@{}c@{}}HumanEval vs\\ Real-World \\ Projects\end{tabular}}} & \multicolumn{1}{c}{\textbf{\begin{tabular}[c]{@{}c@{}}MBPP vs\\ Real-World\\ Projects\end{tabular}}} \\ \midrule

Operators                                                                                                       & 28.33                                                                                          & 29.13                                                                                   & 29.06                                                                                                                                & 40.21                                                     & 36.89                                                & 11.15 ($\uparrow$)                                                                                                                               & 7.83 ($\uparrow$)                                                                                                                        \\
Variables                                                                                                       & 21.84                                                                                          & 22.27                                                                                   & 21.89                                                                                                                                & 26.30                                                     & 26.35                                                & 4.41 ($\uparrow$)                                                                                                                                & 4.46 ($\uparrow$)                                                                                                                          \\
Function                                                                                                        & 10.60                                                                                          & 11.14                                                                                   & 10.73                                                                                                                                & 4.02                                                      & 18.14                                                & -6.72 ($\downarrow$)                                                                                                                              & 7.41 ($\uparrow$)                                                                                                                        \\
Condition                                                                                                       & 9.98                                                                                           & 10.66                                                                                   & 10.20                                                                                                                                & 13.57                                                     & 7.73                                                 & 3.36 ($\uparrow$)                                                                                                                               & -2.47 ($\downarrow$)                                                                                                                       \\
\begin{tabular}[c]{@{}l@{}}Object Oriented \\ Programming\end{tabular}                                          & 8.40                                                                                           & 7.26                                                                                    & 8.17                                                                                                                                 & 0.00                                                      & 0.00                                                 & -8.17  ($\downarrow$)                                                                                                                             & -8.17 ($\downarrow$)                                                                                                                       \\
Loop                                                                                                            & 5.49                                                                                           & 4.95                                                                                    & 5.25                                                                                                                                 & 8.13                                                      & 6.39                                                 & 2.88 ($\uparrow$)                                                                                                                               & 1.14 ($\uparrow$)                                                                                                                        \\
Exception Handling                                                                                              & 4.17                                                                                           & 3.58                                                                                    & 4.02                                                                                                                                 & 0.10                                                      & 0.02                                                 & -3.93  ($\downarrow$)                                                                                                                             & -4.01 ($\downarrow$)                                                                                                                       \\
String Manipulation                                                                                             & 2.30                                                                                           & 2.11                                                                                    & 2.24                                                                                                                                 & 2.35                                                      & 0.84                                                 & 0.12  ($\uparrow$)                                                                                                                              & -1.40 ($\downarrow$)                                                                                                                       \\
Networking                                                                                                      & 1.27                                                                                           & 1.24                                                                                    & 1.26                                                                                                                                 & 0.00                                                      & 0.00                                                 & -1.26 ($\downarrow$)                                                                                                                              & -1.26 ($\downarrow$)                                                                                                                       \\
Data Structure                                                                                                  & 1.19                                                                                           & 1.28                                                                                    & 1.20                                                                                                                                 & 4.55                                                      & 2.83                                                 & 3.35 ($\uparrow$)                                                                                                                               & 1.63 ($\uparrow$)                                                                                                                        \\
File Handling                                                                                                   & 0.73                                                                                           & 1.41                                                                                    & 0.89                                                                                                                                 & 0.00                                                      & 0.00                                                 & -0.89 ($\downarrow$)                                                                                                                              & -0.89 ($\downarrow$)                                                                                                                       \\
Decorators                                                                                                      & 0.80                                                                                           & 0.46                                                                                    & 0.75                                                                                                                                 & 0.00                                                      & 0.00                                                 & -0.75  ($\downarrow$)                                                                                                                             & -0.75 ($\downarrow$)                                                                                                                       \\
Database                                                                                                        & 0.34                                                                                           & 0.73                                                                                    & 0.43                                                                                                                                 & 0.00                                                      & 0.00                                                 & -0.43  ($\downarrow$)                                                                                                                             & -0.43 ($\downarrow$)                                                                                                                       \\
Concurrency                                                                                                     & 0.38                                                                                           & 0.48                                                                                    & 0.41                                                                                                                                 & 0.00                                                      & 0.00                                                 & -0.41 ($\downarrow$)                                                                                                                              & -0.41 ($\downarrow$)                                                                                                                       \\
Serialization                                                                                                   & 0.49                                                                                           & 0.19                                                                                    & 0.34                                                                                                                                 & 0.00                                                      & 0.00                                                 & -0.34 ($\downarrow$)                                                                                                                              & -0.34 ($\downarrow$)                                                                                                                       \\
Context Managers                                                                                                & 0.28                                                                                           & 0.18                                                                                    & 0.25                                                                                                                                 & 0.00                                                      & 0.00                                                 & -0.25 ($\downarrow$)                                                                                                                              & -0.25 ($\downarrow$)                                                                                                                       \\
\begin{tabular}[c]{@{}l@{}}List Comprehension\end{tabular}                                                   & 0.23                                                                                           & 0.21                                                                                    & 0.22                                                                                                                                 & 0.39                                                      & 0.42                                                 & 0.17 ($\uparrow$)                                                                                                                               & 0.20  ($\uparrow$)                                                                                                                       \\
\begin{tabular}[c]{@{}l@{}}Anonymous Function\end{tabular}                                                   & 0.09                                                                                           & 0.03                                                                                    & 0.08                                                                                                                                 & 0.39                                                      & 0.39                                                 & 0.31 ($\uparrow$)                                                                                                                               & 0.31 ($\uparrow$)                                                                                                                        \\
Closures                                                                                                        & 0.07                                                                                           & 0.02                                                                                    & 0.05                                                                                                                                 & 0.00                                                      & 0.00                                                 & -0.05 ($\downarrow$)                                                                                                                              & -0.05 ($\downarrow$)                                                                                                                       \\
Generators                                                                                                        & 0.07                                                                                           & 0.02                                                                                    & 0.05                                                                                                                                 & 0.00                                                      & 0.00                                                 & -0.05  ($\downarrow$)                                                                                                                             & -0.05 ($\downarrow$)                                                                                                                       \\ \bottomrule
\end{tabular}
    }
    \end{minipage}
}
\end{table}

The comparison reveals notable differences in the distribution of KU coverage between real-world projects and the benchmark datasets (HumanEval and MBPP). For example, Operators and Variables are the most frequently occurring KUs in all datasets, but they are more dominant in the benchmarks compared to real-world projects. These KUs exhibit the highest positive differences. In HumanEval, the KU coverage of Operators is 40.21\% compared to 29.06\% in real-world projects, a difference of +11.15. Similarly, Variables KU account for 26.30\% in HumanEval versus 21.89\% in real-world projects, resulting in a difference of +4.41. In contrast, several fundamental and important  KUs are significantly underrepresented or entirely absent in the benchmark datasets. Exception Handling, appearing in 4.02\% of KU instances in real-world projects, is scarcely present in HumanEval (0.10\%) and MBPP (0.02\%), showing absolute differences of -3.93\% and -4.01\%, respectively. A negative difference indicates that a benchmark has a lower KU coverage than real-world projects. Similarly, Object-Oriented Programming, which constitutes 8.17\% of KU instances in real-world projects, is missing in both HumanEval and MBPP. Other advanced KUs such as File Handling, Serialization, Database, and Concurrency which appear with less than 1\% frequency in real-world projects are also absent in both benchmarks.

In addition, we compare the KU coverage of real-world projects with that of each benchmark under study. We observe that the KU coverage of real-world projects is statistically different from both the HumanEval and MBPP datasets.

These findings show that although benchmarks like HumanEval and MBPP capture basic programming KUs, their KU distributions are not representative of real-world Python projects—because the proportional patterns of KU usage in benchmarks differ from those observed in real-world projects. This imbalance suggests that current benchmarks may not fully evaluate the code-generation capabilities of LLMs across the full breadth of the Python’s conceptual space (i.e., KU space), reinforcing the need for KU-aware benchmarks that more accurately reflect real-world programming practices.


\subsubsection{{RQ1.2: \RQOneSubTwo}}

\vspace{0.2cm}
\noindent{\textbf{Motivation:}} The validity of benchmark datasets used for evaluating large language models (LLMs) in code generation tasks critically depends on how well they reflect the diversity and distribution of programming language KUs (which represent core programming concepts) encountered in real-world software development. Drawing on fairness theory, particularly the notion of distributive fairness~\citep{german2018my} (fairness in how outcomes are distributed not necessarily equally but appropriately), we posit that real-world projects should serve as a reference model for a ``correct'' distribution of KUs: although the projects may not distribute KUs uniformly, their distribution is grounded in practical use. Therefore, a benchmark’s KU distribution is more credible to the extent that it demonstrates equitable coverage of KUs -- meaning that the closer a benchmark’s KU distribution aligns with that of real-world software, the more accurate' and representative the benchmark data can be considered in terms of KUs. Our study investigates how closely benchmark datasets align with real-world projects' KU distributions and identifies potential disparities—where over-or-under-representation of certain KUs may bias the evaluation of LLMs, leading to results that do not generalize well to real-world software development.

\vspace{0.2cm}
\noindent{\textbf{Approach:}} To measure the disparity of a KU distribution within benchmarks and real-world projects, we generate the Lorenz curve~\citep{lorenz1905,atkinson1970measurement} and calculate the Gini index~\citep{gini_index}. The Lorenz curve and Gini index are designed to measure inequality of income in a population~\citep{atkinson1970measurement}. The Lorenz curve is a graphical tool that shows how evenly a variable, like income, is distributed among a population. The shape of the Lorenz curve reflects the degree of inequality within the distribution. When the Lorenz curve is a straight line (i.e., 45 degree line), it represents perfect equality, indicating an even distribution of the variable across the population. As inequality increases, the Lorenz curve bends away from this straight line, reflecting greater disparity in the distribution.  To measure the dispersion numerically, we calculate the Gini index~\citep{gini_index}. The Gini index is calculated based on the area between the Lorenz curve and the line of perfect equality, relative to the total area under the line of equality. This index ranges from 0.0 to 1.0, with 0.0 representing perfect equality and 1.0 representing perfect inequality. A Gini index close to one indicates a greater inequality (e.g., the distribution is more skewed).

\vspace{0.2cm}
\noindent{\textbf{Findings:}} 
   
\begin{figure}[!h]
	\centering
	\includegraphics[width=1\linewidth]{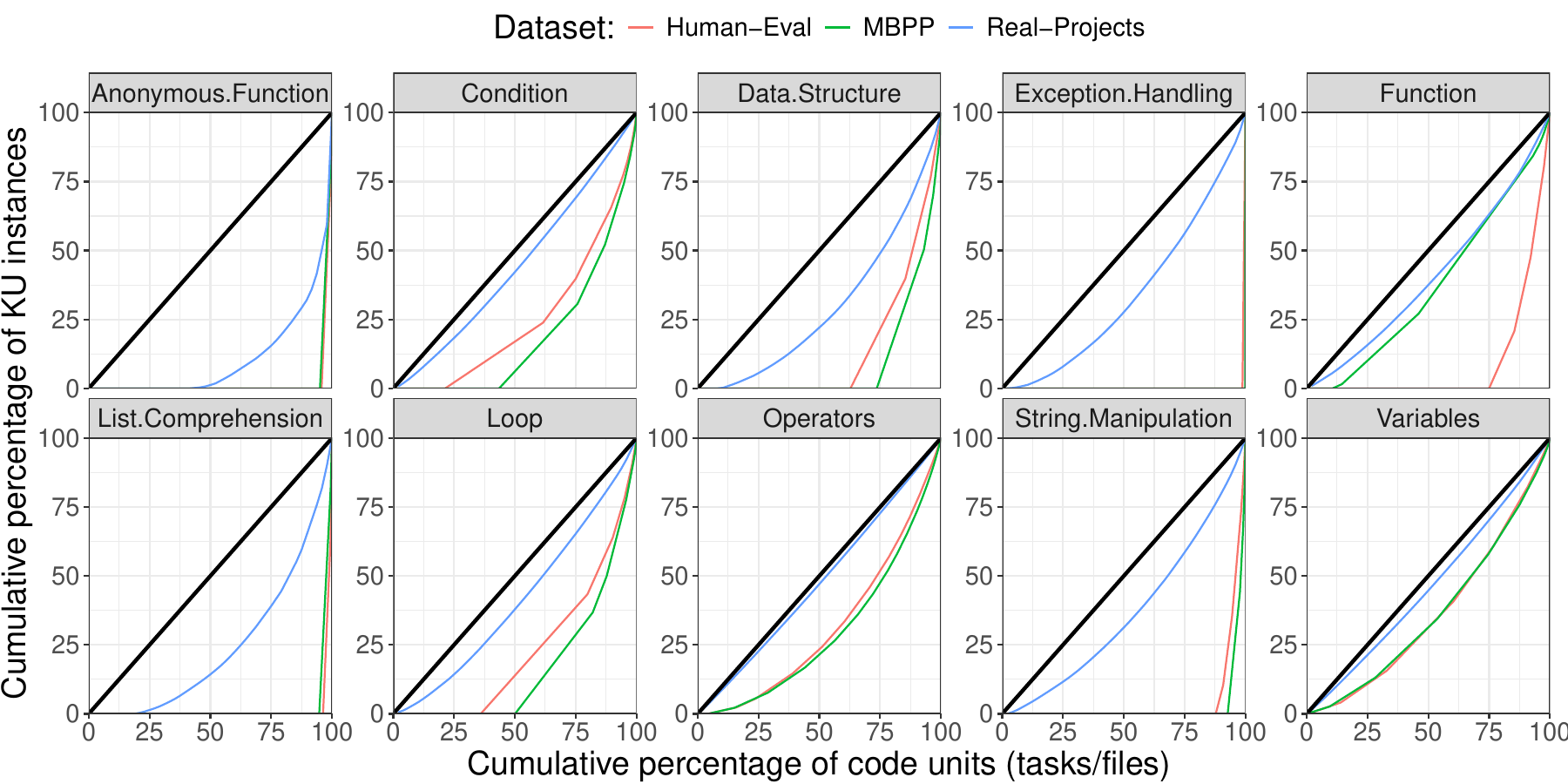}
	\caption{The Lorenz curves of real-world projects and benchmarks that are present in both datasets.} 
    \label{fig:comp_lorenz_real_bench}
\end{figure}

\smallskip \observation{Benchmark datasets exhibit skewed coverage of many fundamental and advanced Python KUs, while real-world projects demonstrate more balanced and comprehensive distributions.}  Figure~\ref{fig:comp_lorenz_real_bench} presents the Lorenz curves for the KUs that are present both in benchmarks and real-world projects, allowing for a direct comparison of their KU distribution among different datasets (e.g., real-world projects and benchmarks). 

The Lorenz curves for HumanEval (red) and MBPP (green) lie below the curves for real-world projects (blue) across almost all KUs, indicating greater inequality in KU distribution. This implies that in benchmarks, a small fraction of tasks accounts for the majority of KU instances, while real-world projects exhibit a more uniform and realistic spread. For KUs such as Exception Handling, List Comprehension, String Manipulation and Anonymous Function, the benchmark curves rise very late and sharply, touching the x-axis before a steep incline. This pattern reveals that these KUs in the benchmarks are either rare or extremely concentrated in only a handful of tasks. In contrast, Lorenz curves of real-world projects for these KUs show a broader distribution, indicating more natural usage patterns. String Manipulation and Exception Handling KUs have Gini coefficient values of 0.27 and 0.32 respectively, in real-world projects, compared to values exceeding 0.85 in both benchmark datasets, highlighting a strong inequality in benchmark KU coverage. In KUs like Operators, Variables, Condition, and Loop, the real-world Lorenz curves are consistently closer to the 45-degree line of perfect equality, suggesting that real-world projects maintain more even KU usage. Although the Lorenz curves of these KUs in benchmarks are close to the 45-degree line, their curves still lie below those of real-world projects. This suggests that even in well-represented KUs, benchmarks exhibit slightly more concentration, meaning a smaller subset of tasks carries a disproportionately higher number of KU instances compared to real-world projects.  

In conclusion, the higher inequality reflected in the Lorenz curves demonstrates that KU distributions of benchmarks lack distributive fairness relative to real-world software projects. Benchmarks disproportionately emphasize certain KUs while marginalizing others, resulting in distributions that do not reflect the balanced and practice-driven patterns observed in real-world Python codebases.

\begin{footnotesize}
    \begin{mybox}{Summary}
    	\textbf{RQ1: \RQOne}
        \tcblower
    	Neither benchmark provides comprehensive coverage of the key KUs of the Python programming language that are commonly used in real-world software projects. In particular,
        
        - [RQ1.1] Only 50\% of the 20 identified KUs are present in each of the studied benchmark datasets individually, whereas real-world projects make use of all the identified KUs. Many of these missing KUs are fundamental to Python programming, including Object-Oriented Programming, Generators, and Context Managers KUs. This result indicates that the KU coverage of benchmarks is not representative of the full conceptual space used in real-world projects.
        
        - [RQ1.2] Benchmark datasets exhibit unequal and disproportionately skewed distributions of the KUs they do include, whereas real-world projects demonstrate a more balanced and comprehensive use of these KUs. This disparity indicates that KU distributions of benchmarks lack distributive fairness relative to real-world software projects.

    
    \end{mybox}
\end{footnotesize}

\subsection{RQ2: \RQTwo}
\label{sec:rq2}

\noindent{\textbf{Motivation:}} In RQ1, we identified a significant KU coverage gap between benchmark datasets and real-world software projects. Many fundamental and advanced Python KUs -- such as Object-Oriented Programming and Exception Handling -- are either missing or severely underrepresented in HumanEval and MBPP benchmark datasets. This imbalance poses a critical limitation: when KUs are absent or underrepresented in benchmarks, they fail to capture the full spectrum of programming concepts that LLMs must master, thereby providing an incomplete and potentially misleading assessment of their true code generation capabilities. In other words, benchmarks may risk overestimating performance on frequently represented KUs while overlooking weaknesses in underrepresented ones, compromising~\citep{german2018my} both the fairness and generalizability of LLM evaluation. To address this limitation, it is essential to develop an approach that can automatically generate KU-based synthesized tasks to improve the overall KU coverage of existing benchmarks, aligning them closer to the KU distribution observed in real-world projects. Enhancing the coverage and diversity of KUs in benchmarks is critical to ensure a more accurate and representative evaluation of LLMs for code generation tasks. Below, we discuss an LLM-based approach that we used to automatically generate KU-based tasks.

\begin{figure}[!h]
	\centering
	\includegraphics[width=1.0\linewidth]{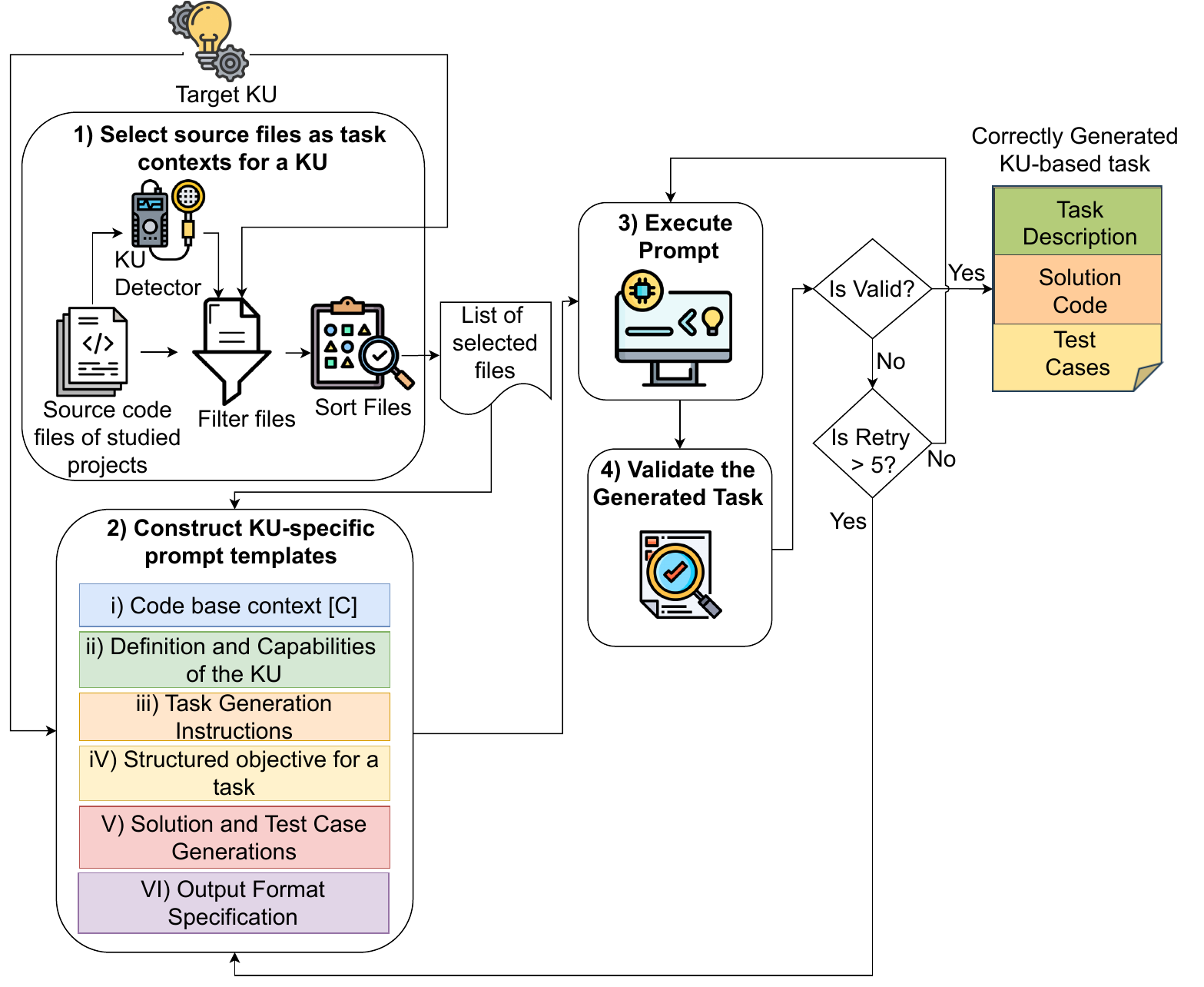}
	\caption{Our LLM-based approach to automatically synthesize KU-based tasks} 
    \label{fig:llm_based_approach_ku_task}
    \end{figure}

\noindent\textbf{Develop an LLM-based approach to automatically generate KU-based tasks.} We design an LLM-based approach that automatically generates KU-based tasks, corresponding solution code for these tasks, and test cases to evaluate the correctness of the generated solutions in the context of code generation. To address the underrepresentation of certain KUs, we focus on generating tasks for those KUs that are missing or underrepresented in the studied  benchmark datasets. Based on our findings in RQ1, we identify 11 such KUs in the two studied benchmarks: Object-Oriented Programming, Exception Handling, Networking, File Handling, Decorators, Database, Concurrency, Serialization, Context Manager, Generators, and Closures. While these KUs are specific to the analyzed benchmarks, our approach is generalizable and can be applied to other benchmarks by targeting a different set of underrepresented or missing KUs. Figure~\ref{fig:llm_based_approach_ku_task} illustrates the overall design of our approach. Below, we outline the key steps involved in designing our LLM-based approach.


\begin{itemize}[itemsep=3pt, wide = 0pt, topsep=1pt, label=$\bullet$, listparindent=\parindent]
	
    \item \textit{Step 1) Select source files as task contexts for each KU.} 
    Our approach begins by selecting a specific KU—referred to as the target KU—for which we aim to generate new tasks. To generate meaningful tasks that are both relevant to a specific KU and applicable to real-world software development, we begin by selecting relevant source files (i.e., \texttt{.py} files) from the studied real-world Python projects. For each KU, we first filter the files to include only those with at least two instances of the targeted KU, ensuring the presence of the KU in those files. We then sort these files in descending order based on their KU instance counts. Files with more instances of the target KU are preferred because they capture richer and more realistic usage patterns, allowing the generated tasks to better reflect how the KU is applied in practice. From this ranked list, we select source files incrementally for task generation. Each task is generated using a distinct source file from the top of the list that serves as the contextual foundation for generating the task corresponding to the target KU. 

    \item \textit{Step 2) Construct KU-specific prompt templates.} To guide the model in generating meaningful, well-structured tasks that exercise a specific KU, we construct KU-specific prompt templates. Each prompt is designed to instruct the model to generate code generation tasks that (i) explicitly require the targeted KU to solve, and (ii) resemble tasks that are encountered in real-world software development. To generate the prompt template, we follow guidelines and examples of prior works~\citep{patel2025get, liu2023your, zhuo2024bigcodebench}. Figure~\ref{fig:llm_prompt_task_generation} in Appendix~\ref{appendix:prompt_tamplate} presents our prompt template for our KU-based task generation.

    \noindent i) \textit{Codebase Context [C]:} A core component of the prompt template is the codebase context \texttt{[C]}. Each codebase context \texttt{[C]} is derived from the ranked list of source files for every KU selected in Step 1. For each task generation for the target KU, we use a distinct source file from the top of the list, ensuring that the generated task is meaningfully grounded in real-world examples of the target KU. This context serves as a foundational reference to help the model generate realistic and appropriate tasks. 

    \noindent \textit{ii) Definition and Capabilities of the KU:} The prompt explicitly specifies the target KU (e.g., Exception Handling) and outlines its core capabilities (cf., Table~\ref{tab:ku-capabilities}). For Exception Handling, these capabilities include: handling exceptions or errors using try/except blocks, raising exceptions using raise, and using finally for guaranteed cleanup actions. This section ensures the model is aware of key programming language concepts it should incorporate when generating tasks related to the specific KU.

    \noindent iii) \textit{Task Generation Instructions:} We provide a set of instructions in the prompt to ensure the model generates tasks that require the target KU. We clearly mention to the model that each task must be testable, with clear input parameters and expected output. The model should use at most three functions from the provided codebase to generate the task ensuring that the task should not be trivial~\citep{crupi2025effectiveness}. To ensure each task is self-contained and easily testable, we mention using Python basic data types to define the task and avoid using project-specific class objects~\citep{crupi2025effectiveness}. In addition, tasks that use only basic data types can be automatically tested across multiple systems without requiring special setup (e.g., defining project-specific classes or importing third-party libraries). 
    
    \noindent \textit{iv) Structured Objective for a Task:} Each task must include a well-defined objective consisting of at least 6–8 clearly specified sub-goals. These sub-goals guide the structure and logic of the solution, ensuring that the use of the target KU is central to the task. The sub-goals must avoid mentioning specific built-in functions; instead, they should describe data flow and return conditions using meaningful variable names and clear natural language. 

    It is important to note that while each synthesized task is designed to emphasize a specific target KU, the implementation of that KU often involves the use of additional supporting KUs (e.g., operators, variables, loops, or condition KUs). This overlap is natural and reflects how KUs interact in real-world programming scenarios. However, since each synthesized task is explicitly centered around the target KU, the occurrence of that KU increases at a much faster rate than the supporting KUs. In contrast, supporting KUs appear only incidentally and therefore grow more slowly in frequency. As a result, the target KU converges faster to the distributions observed in real-world projects than other KUs, helping reduce the representativeness gap while still maintaining realistic interactions among KUs.

    \noindent \textit{v) Solution and Test Case Generation:} The model is further instructed to generate a complete, efficient, and well-structured solution for each task, adhering to best programming practices. Additionally, each task must include five test cases that specify inputs and expected outputs, covering both normal and edge cases to allow thorough validation. To maintain contextual coherence, we do not use a separate prompt for generating the solution and test cases. Instead, they are generated alongside the task within this single prompt, ensuring that both the solution and the test cases are closely aligned with the task’s context and intent.
    
    \noindent \textit{vi) Output Format Specification:} The final part of the prompt specifies a strict JSON format for the output. This format includes fields for the task name, function signature, task description, structured objective, solution code, and test cases. To guide the model in producing consistent and well-structured outputs, we manually generate one example task for each KU, including its detailed task description, structured objective, complete solution, and corresponding test cases. These manually crafted examples serve as output format specifications that illustrate how each component should be organized within the JSON structure. By providing such an example JSON output, we guide the model on how to structure its response consistently~\citep{macedo2024exploring}, ensuring all required components are present and enabling automatic parsing and validation. 

    \item \textit{Step 3) Execute prompts.} We use the \texttt{gpt-4o-mini-2024-07-18} variant of the GPT–4o–Mini model to execute every prompt template and generate tasks corresponding to each KU. We set the temperature parameter to 0.5, striking a balance between creativity and determinism in the responses of the model~\citep{zhang2025thinking,jain2023improving}. We store the responses of the model as JSON objects to further validate in the next step. Each response consists of three components: (i) a task description, (ii) the solution code for the task, and (iii) a set of test cases, each paired with its expected output, to validate the solution. Next, we validate our generated tasks.

    \item \textit{Step 4) Validate the generated tasks.} This step involves multiple validation processes. As an initial checking, we use GPT-4o as a judge to assess whether the solution code is appropriate for the given task description. Prior work also find effective to use LLMs as judge for code generation tasks~\citep{patel2025get, crupi2025effectiveness, majdinasab2025prism}. The judging prompt includes both the task description and the solution code, and the model provides a binary decision (``Yes'' or ``No''). If the output is ``Yes'' we next validate whether the generated task indeed contains the target KU. We apply our KU detector (c.f., Section~\ref{subsec:knowledge_detection}) to identify the presence of the instances of the targeted KU in the solution. A task is considered valid if its solution code includes any instance of the corresponding KU. Next, we validate whether each task is executable. We use Python’s \texttt{exec} command to check that the solution code runs without raising errors. Finally, we verify correctness by executing the test cases and comparing the actual outputs with those expected by the generated test cases.
    
    If the generated task is validated correctly, we select this task as the correctly generated KU-based task. Otherwise, we return to Step 3, re-execute the prompt, and generate a new candidate task until a valid one is obtained. This retry process is performed up to five times. If none of the five attempts produces a valid task, we move to Step 2, select the next file, and generate a new prompt with a new code base context [C]. 
    
\end{itemize}

We repeat Steps 2, 3, and 4 to generate synthesized tasks for each KU. As our stopping criterion, we define N as the number of successfully generated valid tasks for each KU. We begin by generating $N=5$ valid tasks per KU. After generating N valid tasks per KU, we combine these synthesized tasks with the original benchmarks to assess whether the resulting KU distribution becomes statistically similar (cf.  Section~\ref{sec:RQ1_1}) to that of real-world projects. If the distributions remain statistically different, we conduct additional iterations to generate N new valid tasks per KU and reassess the distribution after each iteration. This iterative process continues until the synthesized tasks combined with the benchmark tasks exhibit a KU distribution that is statistically aligned with real-world usage.

Finally, we end up generating 440 valid tasks for each of the 11 selected KUs, yielding an average of 40 tasks per KU. The first and third authors further validate these tasks manually. For the manual validation, the authors randomly select 110 tasks across the selected KUs (i.e., 10 tasks per KU) and collaboratively review them to ensure that (i) each task accurately represents the intended KU, and (ii) the task description, solution, and test cases are consistent, executable, and conceptually aligned with the KU’s key capabilities. Any discrepancies or ambiguities identified during this review are discussed and resolved through mutual consensus. The authors find that the description of every task is meaningful and aligned with the intended KU that is used to generate the task. Examples of KU-based tasks are added in Appendix~\ref{appendix:ku_baesd_task_examples}.

With the newly generated 440 KU-based tasks in hand, we examine the extent to which augmenting existing benchmarks with these synthesized tasks improves their alignment with the KU distributions observed in real-world projects. In addition, we assess how well LLMs can generate correct solutions for these KU-based tasks. In particular, we investigate this research question through two sub-questions:

\begin{itemize}
\item \textbf{RQ2.1:} \RQTowSubOne
\item \textbf{RQ2.2:} \RQTowSubTwo
\end{itemize}


\vspace{0.4cm}
\subsubsection{RQ2.1: \RQTowSubOne}

\vspace{0.2cm}
\noindent{\textbf{Motivation:}} Using our LLM-based approach, we generate synthesized KU-based tasks for the missing or underrepresented KUs observed in the studied benchmarks. It is essential to understand whether incorporating these synthesized tasks with original benchmarks actually leads to a KU distribution that more closely reflects the KU distribution of real-world software projects. 

\vspace{0.2cm}
\noindent{\textbf{Approach:}} We first construct the augmented benchmarks by combining the newly generated KU-based tasks with the raw benchmarks. Then, we measure the improvement in KU-distribution alignment after benchmark augmentation.

\noindent\textbf{Construct the augmented benchmarks.} 
We combine the 440 newly generated KU-based tasks (spanning 11 KUs) with the original HumanEval and MBPP benchmark datasets to construct two augmented benchmarks, namely Augmented-HumanEval and Augmented-MBPP. Each augmented benchmark retains all tasks from the original dataset while incorporating the newly generated tasks containing tasks from all 20 identified KUs. As a result, the Augmented-HumanEval contains a total of 604 tasks, and the Augmented-MBPP contains 1,414 tasks across 20 KUs, offering broader and more balanced KU coverage for evaluating LLMs’ code generation capabilities.

\noindent\textbf{Measuring the improvement in KU-distribution alignment after benchmark augmentation.} To measure how benchmark augmentation improves the alignment of KU distributions with real-world software, we use the Jensen–Shannon Distance (JSDistance)~\citep{endres2003new, lin2002divergence}. JSDistance is a symmetric, bounded metric [0,1] derived from the Jensen–Shannon Divergence~\citep{lin2002divergence}, and is widely used to quantify the similarity between probability distributions. A lower JSDistance value (closer to 0) indicates greater similarity between the KU coverage of a benchmark and real-world projects, whereas a higher value (closer to 1) indicates greater dissimilarity. We compute JSDistance between the KU distribution of real-world projects and the KU distribution of any given benchmark (e.g., HumanEval, MBPP, Augmented-HumanEval and Augmented-MBPP). Finally, to quantify how much augmentation improves KU distribution alignment, we compute the relative improvement using the following formula:

\[
\Delta_{\text{rel}}
= 
\frac{
\mathrm{JSD}\!\left(KU_{\text{projects}}, KU_{\text{orig}}\right)
- \mathrm{JSD}\!\left(KU_{\text{projects}}, KU_{\text{aug}}\right)
}{
\mathrm{JSD}\!\left(KU_{\text{projects}}, KU_{\text{orig}}\right)
}
\]

Here, $KU\_projects$ denotes the KU distribution of real-world projects, $KU\_orig$ refers to the KU distribution of the original benchmark (e.g., HumanEval), $KU\_aug$ denotes the KU distribution of the augmented benchmark (e.g., Augmented-HumanEval) and $JSD$ refers to the function that measures the Jensen–Shannon Distance between two distributions.

\vspace{0.2cm}
\noindent{\textbf{Findings:}} 
\vspace{0.1cm}

\begin{table}[!b]
\centering
\caption{The KU coverage of the original studied benchmarks and augmented benchmarks. The cells highlighted in green indicate KUs for which new tasks were generated, resulting in increased coverage in the augmented benchmarks compared to the original benchmarks.}
\label{tab:aug_benh_ku_ratio}
\resizebox{\columnwidth}{!}{
\begin{tabular}{@{}lrrrrr@{}}
\toprule
\multicolumn{1}{c}{} &
\multicolumn{5}{c}{\textbf{KU coverage (\%)}} \\
\cmidrule(l){2-6}
\textbf{Knowledge Unit (KU)} & \multicolumn{1}{l}{\textbf{HumanEval}} & \multicolumn{1}{l}{\textbf{MBPP}} & \multicolumn{1}{l}{\textbf{\begin{tabular}[c]{@{}l@{}}Real-World\\ Project\end{tabular}}} & \multicolumn{1}{l}{\textbf{\begin{tabular}[c]{@{}l@{}}Augmented-\\ HumanEval\end{tabular}}} & \multicolumn{1}{l}{\textbf{\begin{tabular}[c]{@{}l@{}}Augmented-\\ MBPP\end{tabular}}} \\ \midrule
Operators                    & 40.21                                  & 36.89                             & 29.06                                                                                     & 30.79                                                                                       & 32.10                                                                                  \\
Variables     & 26.30                                  & 26.35                             & 21.89                                                                                     & 21.96                                                                                       & 24.21                                                                                  \\
Function                     & 4.02                                   & 18.14                             & 10.73                                                                                     & 7.23                                                                                        & 11.60                                                                                  \\
Condition                    & 13.57                                  & 7.73                              & 10.20                                                                                     & 8.88                                                                                        & 7.51                                                                                   \\
Object Oriented Programming  & 0.00                                   & 0.00                              & 8.17                                                                                      & \cellcolor[HTML]{93C47D}8.26                                                                & \cellcolor[HTML]{93C47D}7.24                                                           \\
Loop                         & 8.13                                   & 6.39                              & 5.25                                                                                      & 6.53                                                                                        & 6.24                                                                                   \\
Exception Handling           & 0.10                                   & 0.02                              & 4.02                                                                                      & \cellcolor[HTML]{93C47D}4.52                                                                & \cellcolor[HTML]{93C47D}3.58                                                           \\
String Manipulation          & 2.35                                   & 0.84                              & 2.24                                                                                      & 1.01                                                                                        & 0.73                                                                                   \\
Networking                   & 0.00                                   & 0.00                              & 1.26                                                                                      & \cellcolor[HTML]{93C47D}0.93                                                                & \cellcolor[HTML]{93C47D}0.35                                                           \\
Data Structure               & 4.55                                   & 2.83                              & 1.20                                                                                      & 3.65                                                                                        & 2.95                                                                                   \\
File.handling                & 0.00                                   & 0.00                              & 0.89                                                                                      & \cellcolor[HTML]{93C47D}1.07                                                                & \cellcolor[HTML]{93C47D}0.40                                                           \\
Decorators                   & 0.00                                   & 0.00                              & 0.75                                                                                      & \cellcolor[HTML]{93C47D}0.47                                                                & \cellcolor[HTML]{93C47D}0.18                                                           \\
Database                     & 0.00                                   & 0.00                              & 0.43                                                                                      & \cellcolor[HTML]{93C47D}0.25                                                                & \cellcolor[HTML]{93C47D}0.09                                                           \\
Concurrency                  & 0.00                                   & 0.00                              & 0.41                                                                                      & \cellcolor[HTML]{93C47D}1.01                                                                & \cellcolor[HTML]{93C47D}0.38                                                           \\
Serialization                & 0.00                                   & 0.00                              & 0.34                                                                                      & \cellcolor[HTML]{93C47D}0.39                                                                & \cellcolor[HTML]{93C47D}0.15                                                           \\
Context Managers             & 0.00                                   & 0.00                              & 0.25                                                                                      & \cellcolor[HTML]{93C47D}0.72                                                                & \cellcolor[HTML]{93C47D}0.27                                                           \\
List Comprehension           & 0.39                                   & 0.42                              & 0.22                                                                                      & 0.55                                                                                        & 0.48                                                                                   \\
Generators                   & 0.00                                   & 0.00                              & 0.14                                                                                      & \cellcolor[HTML]{93C47D}0.44                                                                & \cellcolor[HTML]{93C47D}0.17                                                           \\
Anonymous Function           & 0.39                                   & 0.39                              & 0.08                                                                                      & 0.19                                                                                        & 0.32                                                                                   \\
Closures                     & 0.00                                   & 0.00                              & 0.05                                                                                      & \cellcolor[HTML]{93C47D}0.16                                                                & \cellcolor[HTML]{93C47D}0.06                                                           \\ \bottomrule
\end{tabular}
}
\end{table}

\smallskip \observation{The overall KU coverage in the augmented benchmarks improves and more closely aligns with that of real-world projects, with JSDistance decreasing from 0.335 to 0.118 for Augmented-HumanEval and from 0.319 to 0.122 for Augmented-MBPP, reflecting an improvement of over 60\% in distributional alignment compared to the original benchmarks.} Table~\ref{tab:aug_benh_ku_ratio} presents the KU coverage across the augmented benchmarks, the original studied benchmarks, and real-world projects. We observe that for KUs that were missing or underrepresented in the original benchmarks (HumanEval and MBPP), the coverage notably increases in the augmented benchmarks (Augmented-HumanEval and Augmented-MBPP). The green-highlighted cells indicate the KUs for which we generate additional tasks, resulting in increased KU coverage after incorporating the newly generated tasks. For instance, the KU coverage for the Object Oriented Programming increases to 8.26\% in Augmented-HumanEval and 7.24\% in Augmented-MBPP, whereas it was not present in the original HumanEval and MBPP datasets. 




In addition, augmenting the benchmarks with KU-based tasks results in KU coverage that more closely aligns with that of real-world projects. The over-representation of certain basic KUs (e.g., Operators and Variables) has been reduced in the augmented benchmarks, bringing their distribution closer to real-world projects. For instance, the KU coverage of Operators decreases from 40.21\% in HumanEval to 30.79\% in Augmented-HumanEval, closely aligning with the 29.06\% observed in real-world projects. Similarly, the KU coverage of Variables decreases from 26.30\% to 21.96\%, aligning more closely with the real-world value of 21.89\%. Moreover, the KU coverage of several KUs has also increased, helping to narrow the gap with real-world distributions. For example, the KU coverage of Exception Handling increases from 0.10\% to 4.52\% in Augmented-HumanEval and from 0.02\% to 3.58\% in Augmented-MBPP, more closely align with the 4.02\% KU coverage observed in real-world projects. 

To quantitatively assess the improvement in alignment, we compute the Jensen–Shannon Distance (JSDistance) between the KU distributions of each benchmark and real-world projects. The JSDistance between HumanEval and real-world projects decreases from 0.335 to 0.118 (i.e., JSDistance reaches closer to zero) after augmentation (Augmented-HumanEval), corresponding to an improvement of approximately 65\% in distributional alignment. Similarly, the JSDistance between MBPP and real-world projects decreases from 0.319 to 0.122 (i.e., JSDistance reaches closer to zero) in Augmented-MBPP, representing an improvement of about 62\%. These reductions in JSDistance indicate that the augmented benchmarks achieve KU distributions that are closer to those of real-world software projects, thereby enhancing their representativeness and fairness for evaluating LLM-based code generation. Furthermore, the Wilcoxon signed-rank test between the KU coverage distributions of real-world projects and the augmented benchmarks shows no statistically significant difference, providing additional evidence that the augmented benchmarks closely reflect the KU coverage distribution observed in real-world projects.
\subsubsection{RQ2.2: \RQTowSubTwo}
\vspace{0.2cm}

\noindent{\textbf{Motivation:}} KUs can be seen as a useful lens to analyze LLMs by revealing where models perform well and where they struggle. Understanding such strengths and weaknesses of LLMs in code generation is crucial for researchers to identify areas for improvement and guide the development of more capable LLMs.

\vspace{0.2cm}
\noindent{\textbf{Approach:}} To evaluate the performance of LLMs on code generation for augmented benchmarks, we select seven popular LLMs: LLaMA3~\citep{dubey2024llama3}, Granite3~\citep{mishra2024granite}, StarCoder2~\citep{li2023starcoder}, Mixtral~\citep{jiang2024mixtral}, Gemma3~\citep{team2025gemma}, GPT-3.5-Turbo~\citep{openai_gpt35_turbo} and GPT-4o-Mini~\cite{hurst2024gpt4oMini}. Table~\ref{tab:llm_categorization} presents the summary of the selected models. This set of selected models provides a comprehensive coverage of the current LLM landscape, spanning different parameter scales from 8B to 175B, architectural designs (dense vs. mixture-of-experts) and different licensing models (open-source vs. proprietary). 

\begin{table}[!t]
\centering
\caption{List of Selected LLMs}
\resizebox{\textwidth}{!}{
\begin{tabular}{llll}
\hline
\textbf{Model} & \textbf{Parameter Scale} & \textbf{Architecture} & \textbf{License} \\ \hline
\makecell[l]{Granite3 \\ \textit{(IBM Research)}} & Small (8B) & Dense & Open-source \\ \midrule
\makecell[l]{Gemma3 \\ \textit{(Google DeepMind)}} & Small--Medium (12B) & Dense & Open-source \\ \midrule
\makecell[l]{StarCoder2 \\ \textit{(BigCode)}} & Medium (15B) & Dense  & Open-source \\ \midrule
\makecell[l]{Mixtral \\ \textit{(Mistral AI)}} & Medium--Large (8$\times$7B, 14B active) & MoE & Open-source \\ \midrule 
\makecell[l]{LLaMA3 \\ \textit{(Meta AI)}} & Large (70B) & Dense & Open-source \\ \midrule
\makecell[l]{GPT-3.5-Turbo \\ \textit{(OpenAI)}} & Very Large (-) & Dense & Proprietary \\ \midrule
\makecell[l]{GPT-4o-Mini \\ \textit{(OpenAI)}} & Large (-) & MoE & Proprietary \\ \bottomrule
\end{tabular}
}
\label{tab:llm_categorization}
\end{table}

We use the task descriptions from the augmented benchmarks as prompts and run the selected models to generate code for each task. To enable comparison, we also evaluate the models on the original benchmark tasks (e.g., HumanEval and MBPP), allowing us to analyze differences in performance between the augmented and original benchmarks. To measure the accuracy of a model we use the \textit{pass@k} metric~\citep{chen2021evaluating}, a widely adopted approach for evaluating code generation performance~\citep{zheng2023codegeex,muennighoff2023octopack,li2024deveval,feng2024complexcodeeval,wang2023recode}. Here, \textit{pass@k} estimates the probability that the model will produce at least one correct solution within the first \textit{k} attempts, as determined by executing the corresponding test cases. We compute pass@k for all models on both the augmented and original benchmarks. To measure the performance drop between an original benchmark and its augmented version, we calculate the relative performance drop using the following formula:

\[
{\text{Relative performance drop}}
=
\frac{
\mathrm{pass@k}_{\text{orig}}
-
\mathrm{pass@k}_{\text{aug}}
}{
\mathrm{pass@k}_{\text{orig}}
}
\times 100\%
\]
Here, $pass@k_{orig}$ denotes the pass@k score on the original benchmark, and $pass@k_{aug}$ denotes the pass@k score on the augmented benchmark. We compute this relative performance drop for every selected model. In this study, we report results for $k = 1, 3,$ and $5$, corresponding to pass@1, pass@3, and pass@5 scores for every model. 

To statistically evaluate whether the performance of LLMs drops in the augmented benchmarks compared to the original ones, we perform a paired Wilcoxon signed-rank test between their performance score distributions for each evaluation level (pass@1, pass@3, and pass@5). We also use the Cliff's delta ($d$)~\citep{cliff_delta} effect size measure to quantify the practical difference between the two distributions. We use the following thresholds for interpreting $d$ \citep{Romano06}: \textit{negligible} for $|\delta| \le 0.147$, \textit{small} for $0.147 < |\delta| \leq 0.33$, \textit{medium} for $0.33 < |\delta| \leq 0.474$, and \textit{large} otherwise.



\vspace{0.2cm}
\noindent{\textbf{Findings:}} 
\vspace{0.1cm}

\observation{When new KU-based tasks are added to original benchmarks the performance of all evaluated LLMs consistently declines in augmented benchmarks compared to the original ones -- highlighting that under-evaluated KUs pose significant challenges for LLMs, likely because current benchmarks give limited attention to these KUs.} Table ~\ref{tab:model-performance-humaneval} presents the comparison of the performance of different LLMs across HumanEval, newly generated KU-based tasks (i.e., NewKUTasks) and Augmented-HumanEval (where the NewKUTasks are added to the original HumanEval). For each model, the last row reports the relative performance drop from HumanEval to Augmented-HumanEval. Compared to models' performance on HumanEval, all models show a notable drop when generating code for NewKUTasks, which in turn drives down their overall performance in Augmented-HumanEval. This performance drop ranges from 12.54\% to 44.82\%  across different evaluation levels (pass@1, pass@3 and pass@5). The Wilcoxon signed-rank test confirms that the performance drop between the original benchmark and its augmented version is statistically significant, with a large effect size observed across all evaluation levels.
 



\begin{table}[!t]
\centering
\caption{Comparison of the performance of different LLMs across HumanEval, newly generated KU-based tasks, and Augmented-HumanEval. The last row for each model reports the {\color{red}relative performance drop (percentage)} from HumanEval to Augmented-HumanEval.}
\label{tab:model-performance-humaneval}
\begin{tabular}{@{}llrrr@{}}
\toprule
\textbf{Model}               & \textbf{Task Type}                                                                                     & \textbf{Pass@1}              & \textbf{Pass@3}              & \textbf{Pass@5}              \\ \midrule
                             & HumanEvaL                                                                                               & 0.74                         & 0.83                         & 0.85                         \\
                             & NewKUTasks                                                                                              & 0.52                         & 0.59                         & 0.60                         \\
                             & Augmented-HumanEvaL                                                                                     & 0.60                         & 0.68                         & 0.70                         \\\cmidrule{2-5}
\multirow{-5}{*}{LLaMA3}    & {\color[HTML]{FE0000} \begin{tabular}[c]{@{}l@{}}Relative Performance Drop \\ (HumanEval → Augmented)\end{tabular}} & {\color[HTML]{FE0000} 18.84\%} & {\color[HTML]{FE0000} 18.29\%} & {\color[HTML]{FE0000} 18.08\%} \\ \midrule
                             & HumanEval                                                                                               & 0.59                         & 0.75                         & 0.80                         \\
                             & NewKUTasks                                                                                              & 0.43                         & 0.54                         & 0.58                         \\
                             & Augmented-HumanEval                                                                                     & 0.49                         & 0.62                         & 0.67                         \\ \cmidrule{2-5}
\multirow{-5}{*}{StarCoder2} & {\color[HTML]{FE0000} \begin{tabular}[c]{@{}l@{}}Relative Performance Drop \\ (HumanEval → Augmented)\end{tabular}} & {\color[HTML]{FE0000} 16.93\%} & {\color[HTML]{FE0000} 17.37\%} & {\color[HTML]{FE0000} 17.07\%} \\ \midrule
                             & HumanEval                                                                                               & 0.69                         & 0.80                         & 0.83                         \\
                             & NewKUTasks                                                                                              & 0.47                         & 0.57                         & 0.60                         \\
                             & Augmented-HumanEval                                                                                     & 0.56                         & 0.66                         & 0.69                         \\ \cmidrule{2-5}
\multirow{-5}{*}{Granite3}    & {\color[HTML]{FE0000} \begin{tabular}[c]{@{}l@{}}Relative Performance Drop \\ (HumanEval → Augmented)\end{tabular}} & {\color[HTML]{FE0000} 19.13\%} & {\color[HTML]{FE0000} 17.86\%} & {\color[HTML]{FE0000} 16.80\%} \\ \midrule
                             & HumanEval                                                                                               & 0.39                         & 0.57                         & 0.64                         \\
                             & NewKUTasks                                                                                              & 0.16                         & 0.22                         & 0.26                         \\
                             & Augmented-HumanEval                                                                                     & 0.25                         & 0.36                         & 0.40                         \\ \cmidrule{2-5}
\multirow{-5}{*}{Mixtral}    & {\color[HTML]{FE0000} \begin{tabular}[c]{@{}l@{}}Relative Performance Drop \\ (HumanEval → Augmented)\end{tabular}} & {\color[HTML]{FE0000} 36.44\%} & {\color[HTML]{FE0000} 37.44\%} & {\color[HTML]{FE0000} 37.01\%} \\ \midrule
                             & HumanEval                                                                                               & 0.80                         & 0.80                         & 0.80                         \\
                             & NewKUTasks                                                                                              & 0.59                         & 0.62                         & 0.64                         \\
                             & Augmented-HumanEval                                                                                     & 0.67                         & 0.69                         & 0.70                         \\ \cmidrule{2-5}
\multirow{-5}{*}{Gemma3}      & {\color[HTML]{FE0000} \begin{tabular}[c]{@{}l@{}}Relative Performance Drop \\ (HumanEval → Augmented)\end{tabular}} & {\color[HTML]{FE0000} 16.07\%} & {\color[HTML]{FE0000} 13.81\%} & {\color[HTML]{FE0000} 12.54\%} \\ \midrule
                             & HumanEval                                                                                               & 0.69                         & 0.73                         & 0.74                         \\
                             & NewKUTasks                                                                                              & 0.17                         & 0.20                         & 0.21                         \\
                             & Augmented-HumanEval                                                                                     & 0.38                         & 0.41                         & 0.42                         \\ \cmidrule{2-5}
\multirow{-5}{*}{GPT-3.5-Turbo}    & {\color[HTML]{FE0000} \begin{tabular}[c]{@{}l@{}}Relative Performance Drop \\ (HumanEval → Augmented)\end{tabular}} & {\color[HTML]{FE0000} 44.82\%} & {\color[HTML]{FE0000} 43.86\%} & {\color[HTML]{FE0000} 43.12\%} \\ \midrule 

 & HumanEval                                                                                               & 0.89                         & 0.90                         & 0.91                         \\
                             & NewKUTasks                                                                                              & 0.46                         & 0.53                         & 0.55                         \\
                             & Augmented-HumanEval                                                                                     & 0.63                         & 0.68                         & 0.69                         \\ \cmidrule{2-5}
\multirow{-5}{*}{GPT-4o-Mini}    & {\color[HTML]{FE0000} \begin{tabular}[c]{@{}l@{}}Relative Performance Drop \\ (HumanEval → Augmented)\end{tabular}} & {\color[HTML]{FE0000} 29.69\%} & {\color[HTML]{FE0000} 24.73\%} & {\color[HTML]{FE0000} 23.26\%} \\

\bottomrule 
\end{tabular}
\end{table}

Among the studied models, Gemma3 exhibits the lowest performance drop (12.54–16.07\%), suggesting relatively stronger robustness to KU-based tasks, whereas GPT-3.5-Turbo suffers the largest drop (43.12–44.82\%), indicating its greater difficulty in handling these newly generated KU-based tasks. Other models, such as LLaMA3, Granite3, and StarCoder2, experience moderate declines (around 16–19\%) while Mixtral lies closer to the high end with drops above 36\%. Interestingly, GPT-4o-Mini, although it achieves the highest performance on the original HumanEval across all evaluation levels (around 0.90 in pass@1, pass@3, and pass@5), suffers a notable decline of 23.26\% to 29.69\% in Augmented-HumanEval, where its scores drop to 0.63, 0.68, and 0.69 for pass@1, pass@3, and pass@5, respectively.

In Augmented-MBPP, we also observe statistically significant performance drops across all evaluation levels, with reductions ranging from 4\% to 15\% (see Table~\ref{tab:model-performance-mbpp} in Appendix~\ref{appendix:ku_augment_MBPP}). The corresponding effect sizes span from small to large, indicating that the impact varies by evaluation level (pass@1, pass@3, and pass@5).

These consistent performance drops across diverse LLMs underscore that current benchmarks may insufficiently represent important language capabilities, emphasizing the need for need for comprehensive coverage of KUs to ensure fair and complete evaluation of a model's efficiency.

\begin{figure}[!t]
	\centering
	\includegraphics[width=1.0\linewidth]{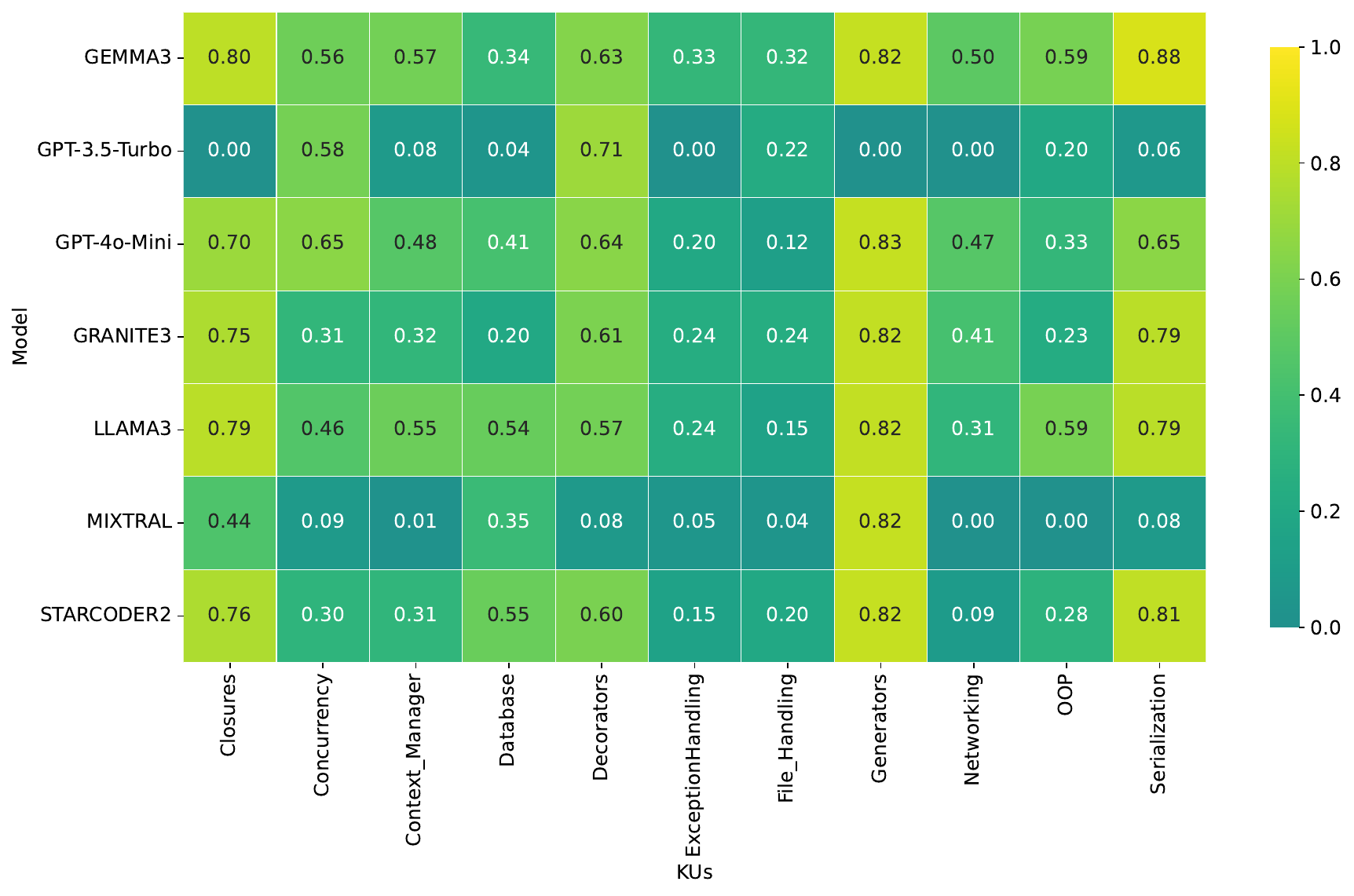}
	\caption{Heat map of the pass@1 of the studied models across KUs for the generated KU-based tasks. The color of each cell encodes the pass@1 value, ranging from green (lower values) to yellow (higher values).} 
    \label{fig:my_task_pass1}
\end{figure}

\observation{Different LLMs show distinct strengths and weaknesses across KUs -- indicating that no single model dominates across all KUs.} Figure~\ref{fig:my_task_pass1} presents the heat map of the performance of the studied models across different KUs for pass@1 for the newly generated KU-based tasks. The heat map color range spans from 0.0 to 1.0 (as shown by the color bar on the right), where darker green indicates lower model performance (values closer to 0) and lighter yellow indicates higher model performance (values closer to 1). 

We observe that certain KUs pose greater challenges for most models. For example, in Exception Handling and File Handling, nearly all models score below 0.24, with Gemma3 performing relatively better (above 0.30). These findings suggest that tasks involving structured control of abnormal conditions and safe input/output operations with resource management remain difficult for LLMs. In contrast, some KUs are handled more effectively across models. For example, in Generators and Closures, several models, including Gemma3, LLaMA3, StarCoder2, and Granite3, score above 0.75, showing that models are comparatively better at capturing stateful iteration (Generators KU) and functional lexical scoping (Closures KU).

Figure~\ref{fig:my_task_pass1} further highlights how different models excel in different KUs. Gemma3 demonstrates steady performance across most KUs but lags in Database, achieving a pass@1 score of 0.34, whereas LLaMA3 and StarCoder2 perform considerably better with scores of 0.54 and 0.55, respectively. In OOP, both Gemma3 and LLaMA3 lead with scores of 0.59, while most other models fall below 0.35. StarCoder2 shows strength in Serialization (0.81) and Database (0.55) but shows its limitations in Networking (0.09). Granite3 maintains moderate scores overall, showing comparatively strong performance in Generators (0.82) and Serialization (0.79), but it does not excel consistently across all KUs. Among the proprietary models, GPT-4o-Mini achieves strong results across multiple KUs excelling in Generators with a pass@1 score of 0.83, but struggling in Exception Handling and File Handling with scores as low as 0.12. At the lower end, GPT-3.5-Turbo and Mixtral achieve near-zero scores in several KUs (e.g., Networking), though they show isolated peaks: GPT-3.5-Turbo in Decorators (0.71) and Mixtral in Generators (0.82). 

These results highlight that each model exhibits a unique profile of KU-specific strengths and weaknesses that reinforces the importance of KU-level evaluation in uncovering nuanced model capabilities that current benchmarks (e.g., HumanEval and MBPP) may fail to fully capture.

\begin{footnotesize}
    \begin{mybox}{Summary}
    	\textbf{RQ2: \RQTwo}
        \tcblower
        Yes, we develop a prompt-based LLM approach that can automatically generate KU-based tasks and augment the KU coverage of benchmark datasets. In particular,

        \begin{itemize}
            \item The augmented benchmarks improve the KU coverage which closely aligns with the KU coverage of real-world projects. After augmentation, the JSDistance between benchmarks and real-world projects drops notably—by about 65\% for HumanEval (0.335$\rightarrow$0.118) and 62\% for MBPP (0.319$\rightarrow$0.122) -- indicating substantially improved distributional alignment.

            \item All studied LLMs show a consistent drop in performance on the augmented benchmarks compared to the original ones. The performance drop between the
            original benchmark and its augmented version is statistically significant, with a large effect size observed in each evaluation level (pass@k, where $k = 1, 3, 5$). In Augmented-HumanEval, this drop ranges from 12.54\% to 44.82\% in all evaluation levels, indicating that LLMs struggle to generate correct code involving the targeted KUs.

             \item No single model dominates across all KUs showing distinct strengths and weaknesses of LLMs' capabilities in code generation  across KUs.
           
        \end{itemize}
        
    \end{mybox}
\end{footnotesize}

	\section{Implications}
\label{sec:Discussion}

In this section, we outline the key implications of our study for both researchers and project maintainers.

\subsection{Implications for researchers}

\textbf{When designing and constructing benchmarks, researchers should ensure that all key KUs are represented so that the benchmark provides comprehensive coverage of programming language concepts.} Our findings in RQ1.1 show that existing popular benchmarks like HumanEval and MBPP omit nearly half of the identified 20 Python KUs, limiting their ability to fully evaluate model performance. Moreover, our analysis in RQ1.2 reveals that even among the KUs present in these benchmarks, the distribution is highly imbalanced — certain basic KUs (e.g., Operators, Control Flow) dominate, while intermediate and advanced KUs receive little representation. This skew further reduces the benchmarks’ ability to reflect real-world programming practices, where projects exhibit a more balanced use of diverse KUs. Comprehensive and balanced KU coverage is essential because code generation tasks inherently span a wide range of programming concepts, from simple operator KU to advanced KUs such as concurrency, database, and exception handling. By ensuring that all key KUs are represented, researchers can create benchmarks that better reflect the diversity of real-world programming practices, provide more accurate assessments of model capabilities, and ultimately guide the development of LLMs that are robust across the full spectrum of programming language features.

\textbf{Our KU-based evaluation can help researchers pinpoint weaknesses of LLMs and guide focused improvements.} KU-based evaluation enables researchers to move beyond aggregate performance scores and uncover the specific strengths and weaknesses of LLMs through the lens of KUs. In RQ2.2, we observe that certain KUs pose greater challenges for most of the studied models. For example, in Exception Handling and File Handling, nearly all studied models pass@1 score below 0.24, with Gemma3 performing relatively better (above 0.30). In addition, we observe that different LLMs have distinct strengths and weaknesses across KUs. We observed that StarCoder2 shows strength in Serialization (0.81) and Database (0.55) but shows limitations in Networking (0.09). By identifying the KUs where a model consistently underperforms, researchers can make better use of their limited resources by directing efforts toward targeted fine-tuning~\citep{zheng2024fine, fan2025fait} or curriculum learning~\citep{ryu2025curricullm, kim2024strategic, chen2025self}, prioritizing those weaker areas to develop more robust and balanced models.


\subsection{Implications for project maintainers}

\textbf{Our framework enables project maintainers to identify the most suitable LLM for their projects.} When project maintainers need to decide which LLM is most suitable for code generation in their projects, they can apply our framework to evaluate models using KU-based tasks derived from their own codebase. As we observe in our study, all the studied LLMs suffered a significant performance drop (12.54–44.82\%) on the augmented benchmark and this performance drop is statistically significant with a large effect size across all evaluation levels (i.e., pass@1, pass@3 and pass@5). To ensure the quality and correctness of these tasks, the authors manually validated each KU-based task to confirm its alignment with the intended KU and to verify that the corresponding solution and test cases were functionally correct. Importantly, each model exhibits unique strengths and weaknesses specific to different KUs. For instance, some LLMs perform better on Generators but poorly on Exception Handling or Database KU tasks. This result suggests that if a project is heavily reliant on database operations, maintainers may select an LLM that demonstrates stronger performance on the Database KU. This KU-aware evaluation provides actionable insights, helping teams to select the LLM that best aligns with their project’s practical needs.


	\section{Related Work}
\label{sec:Related_Work}

In this section, we review related work across three areas: (1) manually curated benchmarks for code generation, (2) synthetic and dynamic benchmarks designed to expand or adapt evaluation tasks, and (3) studies on benchmark quality.

\noindent\textbf{Manually curated benchmarks for code generation tasks:} The most widely used benchmark for evaluating the performance of LLMs in code generation task is HumanEval~\citep{chen2021evaluating}. This benchmark contains 164 code generation tasks that are manually designed by OpenAI researchers. Each task includes a function signature, docstring, body, and several unit tests (on  average seven per task). HumanEval focuses mainly on assessing programming language comprehension, reasoning, algorithms, and simple mathematics. Another popular code generation benchmark is MBPP~\citep{austin2021program}, which comprises of 974 Python programming problems that are designed to be solved by entry-level programmers. Crowd-sourced participants design the benchmark by writing problem statements, solutions, and three test cases to evaluate function-level code completion. Beyond Python, some benchmarks extend HumanEval and MBPP to other languages. For example, HUMANEVAL-X~\citep{zheng2023codegeex} adds problems related to three other programming languages (e.g., $C{++}$, $JavaScript$, and $Go$) to HumanEval, while MultiPL-E~\citep{cassano2023multipl} expands both HumanEval and MBPP to 18 programming languages. Another work augmented the HumanEval and MBPP benchmarks by generating additional test cases, resulting in $HumanEval^{+}$ and $MBPP^{+}$, which provide a more precise evaluation of the functional correctness of LLM-generated code~\citep{liu2023your}.

Recent work has introduced benchmarks that go beyond function-level tasks, targeting more software development settings (e.g., class-level). ClassEval~\citep{du2024evaluating} is the benchmark constructed manually at the class level, consisting of 100 Python tasks that require generating classes. CoderEval~\citep{yu2024codereval} focuses on repository-level code generation by collecting 230 Python and 230 Java tasks from real-world open-source projects. This benchmark evaluates both standalone and non-standalone functions while supporting multilingual code generation. Extending this line of work, DevEval~\citep{li2024deveval} offers a large-scale, manually annotated benchmark with 1,874 testing samples drawn from 117 repositories across 10 popular domains. By including requirements, original repositories, and reference dependencies, DevEval aims to more closely simulate realistic coding processes within working repositories. Along similar lines, HumanEvo~\citep{zheng2024humanevo} incorporates evolution-awareness by rolling back repositories to pre-commit states, providing 400 tasks (200 Python, 200 Java) categorized by dependency levels (standalone, intra-class, inter-class) with both detailed and brief docstrings. Finally, CODEAGENTBENCH~\citep{zhang2024codeagent} complements these efforts by offering 101 repository-level tasks—functions and classes drawn from real Python projects across five domains (machine learning, data structures, information extraction, databases, and networking)—to capture complex, real-world development challenges.

None of the above works study whether these benchmarks cover all key language concepts of programming languages. In this work, we study the representativeness of benchmarks through KUs of programming languages, understand gaps compared to the KUs used in real-world projects, and further provide a framework to dynamically generate KU-based tasks that can augment existing benchmarks in terms of KU coverage.

\noindent\textbf{Synthetic and dynamic benchmark for code generation tasks:} Synthetic benchmarks leverage LLMs to automatically create evaluation data, improving scalability, and diversity. CODEBENCHGEN~\citep{xie2024codebenchgen} demonstrates this approach by adapting real code fragments into executable examples with test cases, resulting in datasets such as Exec-CSN (1,931 examples from 367 GitHub repositories) that emphasize diversity, realism, and solvability. Building on this idea, BigCodeBench~\citep{zhuo2024bigcodebench} integrates human oversight with LLM generation to construct 1,140 rich-context Python tasks involving multi-tool use, aiming to bridge the gap between simple coding exercises and real-world programming. 

More recently, researchers have developed a realistic code synthesis benchmark (SWE-Bench) by pairing GitHub issues with their corresponding codebases and test cases, enabling the evaluation of LLMs on tasks that closely reflect real-world software engineering scenarios~\citep{jimenez2024swebench}. Extending the scope further, ComplexCodeEval~\citep{feng2024complexcodeeval} leverages LLMs to generate high-quality docstrings and collects over 11,000 samples (Java and Python) from high-star GitHub projects, supporting multiple tasks such as code generation, completion, test case generation, and API recommendation while proactively avoiding data leakage. In another work, Prism~\citep{majdinasab2025prism} introduces a dynamic benchmarking framework that adapts to model capabilities and systematically explores difficulties in tasks using Monte Carlo Tree Search (MCTS). The authors formulate the benchmarking process as a search problem with the objective of identifying model’s capabilities and limitations through adaptive exploration of computer science concepts (e.g., dynamic programming, and algorithm) with various difficulty levels (e.g., very easy, easy, medium, hard, and very hard). The generated tasks are similar to the LeetCode programming challenges. 

Our work differs from above works by focusing on key KUs (e.g., concurrency, database and object-oriented programming) of the Python language that are fundamental and important language concepts of the Python. In addition to task diversity, we systematically ensure that all essential language constructs are represented in the evaluation. Furthermore, in generating our benchmark, we leverage real-world code as the foundation for task creation, ensuring stronger relevance to programming challenges.

\noindent\textbf{Benchmark dataset Quality:} Large-scale progress in code generation has leaned heavily on different public benchmark datasets, yet a growing body of work shows that these suites are vulnerable to data leakage and test contamination. \citet{matton2024leakage} systematically document contamination pathways including direct leakage, synthetic-data transfer, and selection overfitting. The authors introduce LBPP as a harder, decontaminated alternative and find that model performance drops noticeably when evaluated on this dataset compared to contaminated ones. \citet{riddell2024quantifying} quantify both surface-level and semantic overlap between popular code benchmarks and pretraining corpora, finding substantial duplication and inflated scores on items similar to training data.  Complementing these leakage-centric findings, \citet{siddiq2024fault} assess prompt quality across nine code benchmarks and report issues (e.g., clarity, grammar, narrow language coverage) that can distort model performance and trustworthiness—alongside evidence of memorization concerns. 

Beyond code generation specifically, several broader studies outline contamination mechanisms and propose detection/mitigation tools. \citet{xu2024benchmark} survey benchmark data contamination (BDC), organizing methods into detection and mitigation families and highlighting risks when training data are opaque. In another work, \citet{zhou2023don} conduct extensive experiments and their result shows that even partial leakage can dramatically boost benchmark results. The authors also offer guidelines for benchmark designers and users. \citet{dong2024generalization} present distribution-based detection (CDD/TED) that flags contamination using only model outputs—useful when training corpora are unavailable. The results show that the contamination detection achieves the average relative improvements of 21.8\%-30.2\% over other contamination detection approaches.

Although prior research has primarily concentrated on issues of data leakage and contamination in benchmark evaluation, relatively little attention has been paid to whether existing code-generation benchmarks comprehensively capture the KUs of programming languages. Our work addresses this gap and study the comprehsiveness of KUs of programming languages in benchmarks. In addition, we design an LLM-based approach to automatically generate tasks for a specific KU to improve KU coverage.



    \section{Threats to Validity}
\label{sec:Limitations_And_Threats}



\smallskip \noindent\textbf{Construct validity.} A potential construct validity threat is that our KU-based task generation framework does not incorporate class or module dependencies. Our primary focus in this work, however, is on programming language features -- captured through KUs such as exception handling, concurrency, generators, or decorators -- rather than on higher-level structural complexity. The former is important because language features represent the fundamental capabilities provided by a programming language, expressed through its constructs and APIs, and form the foundation upon which larger program structures are built. Modeling these language features as KUs offers a principled first step toward making benchmarks more representative, as developers necessarily rely on these features when structuring programs using classes, inheritance, and modules. Thus, although we do not model full architectural dependencies, the language features embodied in KUs capture substantial and meaningful aspects of how real software is written. Future work could investigate generating KU-based tasks that require handling class- or module-level dependencies by treating such dependency-related knowledge as framework/environment KUs, thereby broadening the scope of KUs to cover build-system, dependency-resolution, and module-organization capabilities that play a significant role in real-world software development.

Another construct validity threat arises from how we select source files as contextual input for generating tasks for a KU. In our approach, we sort candidate files in descending order based on their KU instance counts and select files from the top of this list. This prioritization aims to ensure that the selected files contain richer and more realistic usage patterns of the KU, increasing the likelihood that the synthesized tasks reflect how the KU is applied in practice. However, this approach emphasizes the frequency of KU occurrences rather than the specific capabilities exercised within a KU, which may correspond to different difficulty levels. The primary focus of this paper is to investigate whether improving benchmark representativeness influences model performance; analyzing capability-level difficulty and its impact on model behavior is beyond our current scope. Future work could examine KU capabilities across varying difficulty levels, incorporate capability-level weighting schemes, and study model performance across these differences.

\smallskip \noindent \textbf{Internal validity}. A potential internal validity threat arises from our use of a fixed batch size N=5 when generating synthesized tasks for each KU in each iteration for HumanEval and MBPP. The specific choice of N may influence the exact number of tasks ultimately produced for other benchmarks. Different values of N could also lead to faster convergence or yield slightly different synthesized task sets. Our choice of N=5 ensures a small batch size, thereby limiting the excessive invocation of LLMs during task generation and validation. Nonetheless, future work may explore data-driven or optimization-based strategies for determining batch size or stopping criteria to more systematically guide the task-generation process.

\smallskip \noindent \textbf{External validity.} Our study focuses on two widely used benchmarks, HumanEval and MBPP, to analyze KU coverage and understand their gaps compared to real-world projects. While these benchmarks are widely adopted in the literature and provide a strong starting point, there are other manually curated benchmarks or synthetic/dynamic benchmarks. Different benchmarks may emphasize alternative programming concepts, have varying task structures, or exhibit different distributions of KUs. To mitigate this limitation, we provide a supplementary package of our framework that allows researchers to easily apply our KU-based analysis to additional benchmarks. As the first study to systematically examine benchmarks through the lens of KUs, our work establishes a foundation that future researchers can build on to investigate KU coverage across other benchmarks and broaden the understanding of how comprehensively LLM benchmarks reflect real-world programming practices.

Another external validity threat is that our study focuses only on Python KUs. While Python is one of the most widely studied languages in LLM evaluation, our findings may not directly generalize to benchmarks in other programming languages, which differ in syntax, constructs, and KU distributions. Nevertheless, our operationalization of KUs and the LLM-based KU detector are designed to be generalizable, making them applicable to other languages. This provides a pathway for future research to extend KU-based analysis beyond Python, generate benchmarks for other programming languages,  and validate our findings.

    \section{Conclusion}
\label{sec:Conclusion}

\sloppy In this paper, we conducted the first systematic investigation of the representativeness of code generation benchmarks in terms of language concepts modeled as Knowledge Units (KUs) of programming languages, revealing substantial gaps in how existing benchmarks represent the core language concepts used in real-world software development. Our analysis of HumanEval and MBPP—two widely adopted benchmarks—shows that they cover only 50\% the identified 20 Python KUs observed in real-world projects. Moreover, KUs that occur frequently in real-world codebases are either severely underrepresented or entirely missing in the benchmarks. Instead, these benchmarks are heavily skewed in many KUs, whereas real-world projects exhibit a more balanced and comprehensive KU distribution. These disparities demonstrate that the KU distributions of current benchmarks lack representativeness and distributive fairness compared to real-world software projects.

To address these gaps, we developed an LLM-based framework that automatically generates KU-specific tasks. Using this framework, we constructed 440 new tasks across 11 underrepresented KUs in benchmarks, generate augmented benchmarks that substantially improve KU coverage and better align with the KU distribution observed in real-world projects. Evaluating seven popular LLMs on these augmented benchmarks demonstrated a statistically significant performance drop (12.54–44.82\%) with a large effect size, highlighting that under-evaluated KUs pose significant challenges for LLMs, likely because current benchmarks give limited attention
to these KUs. These findings underscore the importance of benchmarks having comprehensive and representative KU coverage to effectively evaluate the performance of LLMs, understand their strengths and weaknesses through the lens of KUs, and ultimately guide the development of more balanced and robust models. 

Overall, our work leveraged KUs as a lens for analyzing benchmark comprehensiveness and evaluating LLMs' performance. As a first step in this area, our study opens avenues for future research to extend KU-based analysis to additional programming languages, incorporate other sources of knowledge units (e.g., domain-specific KUs such as software architecture), and evaluate next-generation agentic LLMs~\citep{plaat2025agentic}. These directions will further advance the reliability and applicability of LLMs in software engineering.



	\section*{Declarations}
\label{sec:declarations}

\subsection*{\textbf{Data Availability Statement (DAS)}}

\noindent A supplementary material package is provided online in the following link:
\url{https://bit.ly/4pt8YEH}. The contents will be made available on a public GitHub
repository once the paper is accepted.

\subsection*{\textbf{Funding and/or Conflicts of interests/Competing interests}} 

\noindent The authors declared that they have no conflict of interest.

	\begin{footnotesize}
		\bibliographystyle{spbasic}      
		\bibliography{bib/references.bib}   
	\end{footnotesize}	

	\clearpage
	
	\appendix

	\begin{Large}
		\noindent \textbf{Appendix}
	\end{Large}
	
	\normalsize
	\vspace{-1ex}

\section{Constructed Prompt Templates} \label{appendix:prompt_tamplate}

\begin{figure}[!h]
    \centering
    \includegraphics[width=1.0\textwidth]{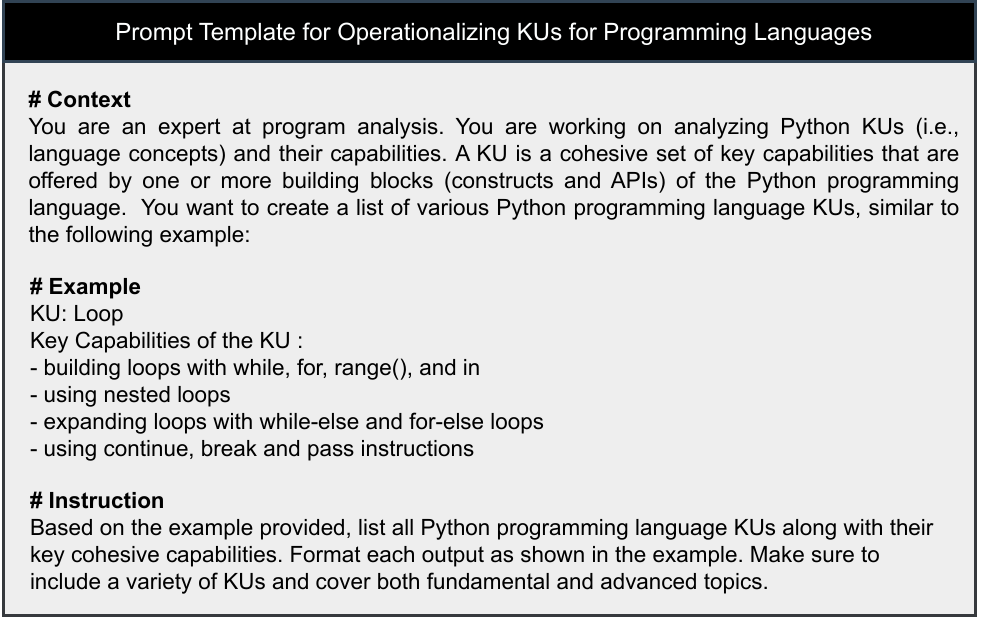}
    \caption{The prompt template to generate the list of Python programming language KUs and their associated key cohesive capabilities.}
    \label{fig:ku_oper_prompt}
\end{figure}

\begin{figure}[!h]
    \centering
    \includegraphics[width=1.0\textwidth]{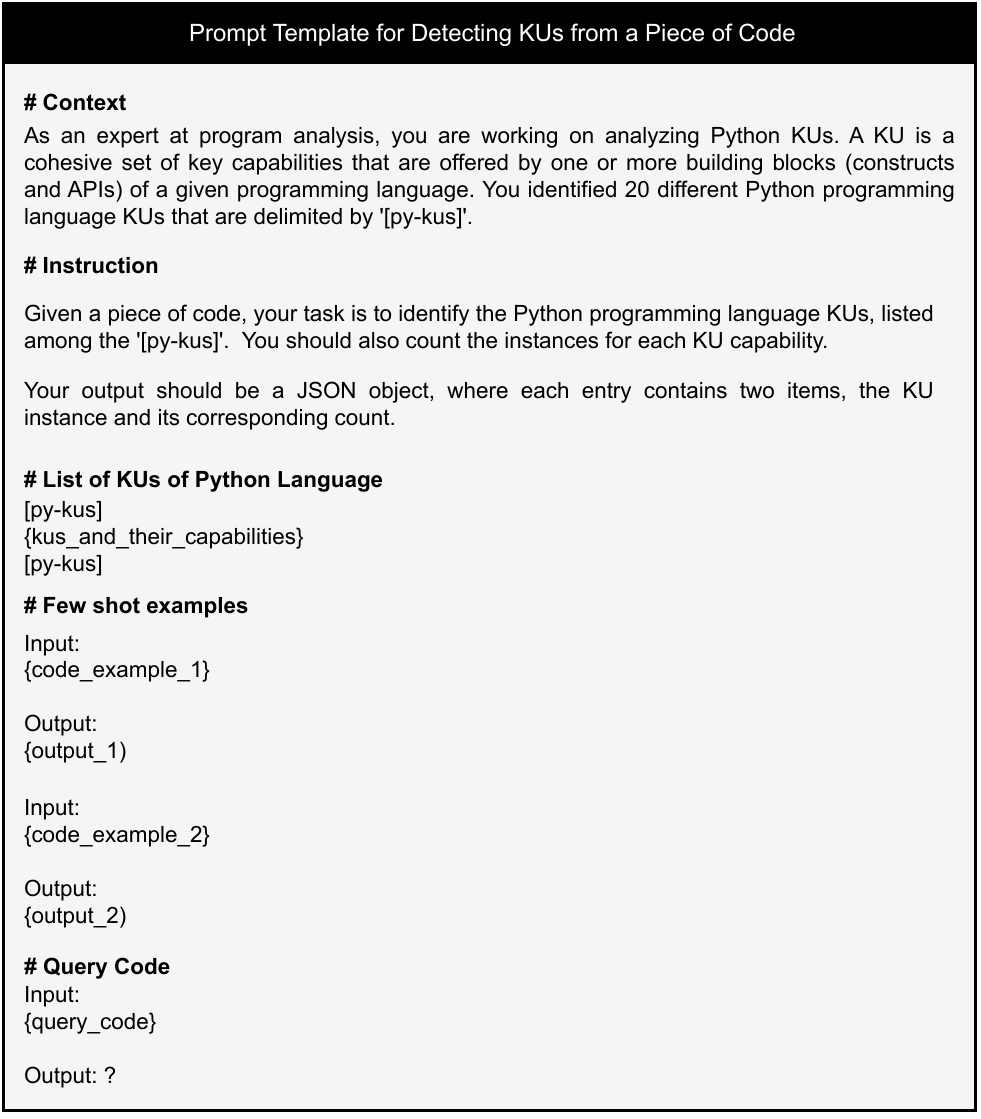}
    \caption{The prompt template to detect the incidence of Python programming language KUs from a given piece of code.}
    \label{fig:ku_detection_prompt}
\end{figure}

\begin{figure}[!h]
	\centering
	\includegraphics[width=1\linewidth]{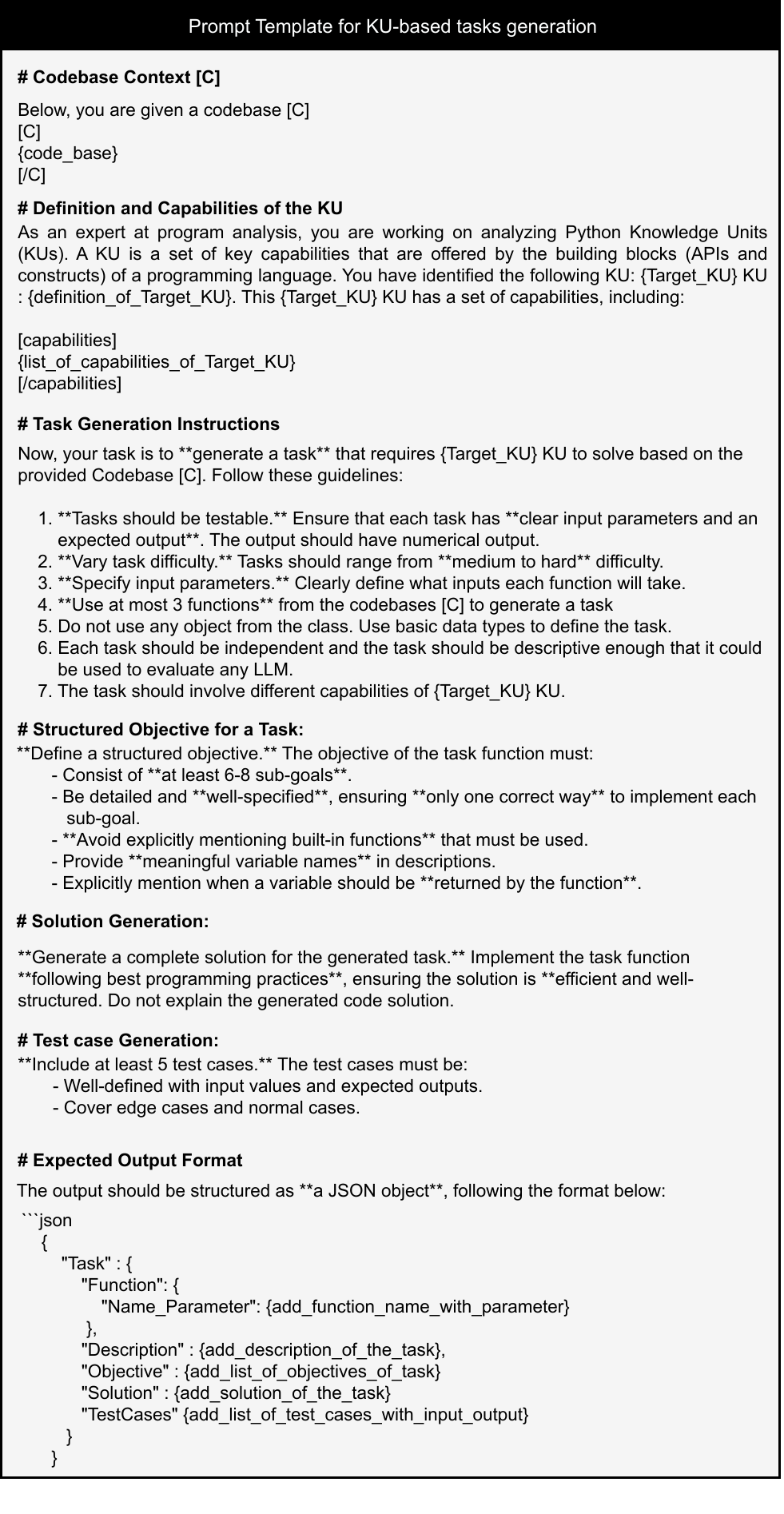}
	\caption{The prompt template for generating KU-based tasks.} 
    \label{fig:llm_prompt_task_generation}
\end{figure}

\FloatBarrier
\section{List of Python Knowledge Units and their Key Capabilities.}\label{appendix:python_kus}

\begin{table}[h!]
    \centering
    \small
    \caption{Python Knowledge Units and their key capabilities}
    \label{tab:ku-capabilities}
    \resizebox{\textwidth}{!}{
    \begin{tabular} {p{3cm} p{10cm}} 
    \toprule
    \multicolumn{1}{c}{\textbf{Knowledge Unit (KU)}} & 
    \multicolumn{1}{c}{\textbf{Key Capabilities}} \\ \midrule
    \textbf{[K1]} Variable & 
    \textbf{[C1]}  Assign variables\\ 
    & \textbf{[C2]}  Use numeric data types (e.g., int, float)\\
    & \textbf{[C3]}   Use string data type\\ 
    & \textbf{[C4]}   Use boolean data type\\ 
    & \textbf{[C5]}  Use \texttt{None} data type\\ \midrule
    
    \textbf{[K2]} Operators & 
    \textbf{[C1]}  Use arithmetic operators: +, -, *, /, //, \% \\ 
    & \textbf{[C2]}  Use comparison operators: <, >, <=, >=, ==, != \\ 
    & \textbf{[C3]}  Use logical operators: and, or, not\\
    & \textbf{[C4]} Use assignment operators: =, +=, -=, *=, /=, //=, \%= \\ 
    & \textbf{[C5]}  Use bitwise operators: \&, |, \textasciicircum, \textasciitilde
, <<, >> \\ 
    & \textbf{[C6]}  Use membership operators: in, not in \\ \midrule

    \textbf{[K3]} Condition & 
    \textbf{[C1]}   Use if, elif, and else statements\\ 
    & \textbf{[C2]}  Use nested conditionals\\ 
    & \textbf{[C3]} Apply short-circuit evaluation \\ 
    & \textbf{[C4]} Use ternary (conditional) expressions \\ \midrule

    \textbf{[K4]} Loop & 
    \textbf{[C1]}   Create while loops\\
    & \textbf{[C2]}   Create for loops using range() and in\\
    & \textbf{[C3]}   Iterate through sequences (e.g., lists, strings)\\
    & \textbf{[C4]}   Use while-else and for-else clauses\\
    & \textbf{[C5]}   Create nested loops\\
    & \textbf{[C6]}   Use pass and continue statements\\ \midrule

    \textbf{[K5]} Function & 
    \textbf{[C1]}   Define functions using def\\
    & \textbf{[C2]}   Use parameters and arguments in function declaration\\
    & \textbf{[C3]}   Return values from the function using return\\
    & \textbf{[C4]}  Access global variables in the function\\
    & \textbf{[C5]}   Use default and keyword arguments \\ \midrule

    \textbf{[K6]} Anonymous Function & 
    \textbf{[C1]}   Use lambda expressions to define anonymous functions \\ \midrule

    \textbf{[K7]} Data Structure & 
    \textbf{[C1]}   Create and manipulate lists\\
    & \textbf{[C2]}   Create and manipulate sets\\
    & \textbf{[C3]}   Create and manipulate dictionaries\\
    & \textbf{[C4]}  Create and manipulate tuples\\
    & \textbf{[C5]}   Append elements to collections \\ 
    & \textbf{[C5]}   Slice and access values in collections \\
    & \textbf{[C5]}   Perform updates and lookups in collections \\\midrule

    \textbf{[K8]} File Handling & 
    \textbf{[C1]}   Open files in various modes (read, write, append, etc.)\\
    & \textbf{[C2]}   Read from files using different functions such as \texttt{read()}, \texttt{readline()} and \texttt{readlines()}\\
    & \textbf{[C3]}   Write to files using different functions such as \texttt{write()} and \texttt{writelines()}\\
    & \textbf{[C4]}  Close files properly using \texttt{close()} \\ \midrule

    \textbf{[K9]} Object-Oriented Programming (OOP)& 
    \textbf{[C1]}   Define classes and create objects\\
    & \textbf{[C2]}   Use instance and class variables\\
    & Define instance methods, overloading methods and access them\\
    & \textbf{[C4]}  Use class methods (@classmethod) \\
    & \textbf{[C5]}  Use static methods (@staticmethod)\\
    & \textbf{[C6]}  Apply inheritance to create subclasses  \\
    & \textbf{[C7]}  Use encapsulation with private/protected members \\
    & \textbf{[C8]}  Apply polymorphism with method overriding \\ \midrule

    \textbf{[K10]} Exception Handling & 
    \textbf{[C1]}   Use \texttt{try} and \texttt{except} blocks to monitor code for exceptions or errors\\
    & \textbf{[C2]} Raise exceptions using the \texttt{raise} statement\\
    & \textbf{[C3]} Handle multiple exceptions\\
    & \textbf{[C4]}  Use \texttt{finally} block for cleanup after handling exception \\ \midrule

    \end{tabular}
    }
\end{table}

\begin{table}[!b]
    \centering
    \resizebox{\textwidth}{!}{
    \begin{tabular} {p{3cm} p{10cm}} 
    \toprule
    \multicolumn{1}{c}{\textbf{Knowledge Unit (KU)}} & 
    \multicolumn{1}{c}{\textbf{Key Capabilities}} \\ \midrule
     \textbf{[K11]} Generators & 
    \textbf{[C1]}   Create generators using the \texttt{yield} statement\\
    & \textbf{[C2]} Iterate over generator objects \\
    & \textbf{[C3]} Use iter() and next() with iterators\\ \midrule
    \textbf{[K12]} Decorators & 
    \textbf{[C1]}   Use function decorators\\
    & \textbf{[C2]} Create decorators with arguments \\
    & \textbf{[C3]} Decorate class methods\\ 
    & \textbf{[C4]} Use built-in decorators: \texttt{@staticmethod}, \texttt{@classmethod}, \texttt{@property} \\ \midrule 
    
    \textbf{[K13]} Closures & 
    \textbf{[C1]}   Define a nested function within an outer function\\
    & \textbf{[C2]} Reference outer function variables in the inner functions \\
    & \textbf{[C3]} Return the inner function to create a closure\\ \midrule
    
    \textbf{[K14]} Context Managers & 
    \textbf{[C1]}   Use the \texttt{with} statement for resource management\\
    & \textbf{[C2]} Create custom context managers using \texttt{\_\_enter\_\_} and \texttt{\_\_exit\_\_} \\
    & \textbf{[C3]} Use the \texttt{contextlib} module for context management \\ \midrule
    
    \textbf{[K15]} Comprehension & 
    \textbf{[C1]}  Create lists using list comprehensions with iteration and optional filtering\\ 
    & \textbf{[C2]}  Construct dictionaries using dictionary comprehensions with key–value pairs\\
    & \textbf{[C3]} Build sets using set comprehensions ensuring unique elements\\ 
    & \textbf{[C4]} Generate sequences lazily using generator expressions\\ 
    & \textbf{[C5]}  Implement nested comprehensions for multi-level iteration and transformation\\ \midrule

    \textbf{[K16]} Concurrency & 
    \textbf{[C1]}  Manage concurrent tasks using the threading module to run multiple threads within a single process\\ 
    & \textbf{[C2]}  Achieve parallel execution with the multiprocessing module by creating separate processes\\
    & \textbf{[C3]} Write asynchronous code using async and await keywords for cooperative multitasking\\ 
    & \textbf{[C4]} Use the concurrent.futures module to simplify concurrent task execution\\ 
    \midrule

    \textbf{[K17]} String Manipulation & 
    \textbf{[C1]}  Create and combine strings using string literals, concatenation (+), and repetition (*)\\ 
    & \textbf{[C2]} Transform case with methods such as upper(), lower(), title(), and capitalize()\\
    & \textbf{[C3]} Clean and trim strings using methods like strip(), lstrip(), and rstrip()\\ 
    & \textbf{[C4]} split and join strings with split(), rsplit(), and join()\\ 
    & \textbf{[C5]}  Modify content with replace() and search with find(), index(), startswith(), and endswith()\\ 
    & \textbf{[C6]} Format strings using format(), f-strings (formatted string literals), and the \% operator\\ 
    & \textbf{[C7]} Search, match, and parse patterns in strings using the re (regular expressions) module\\ 
    \midrule

    \textbf{[K18]} Networking & 
    \textbf{[C1]}  Establish low-level communication using sockets for sending and receiving data over networks\\ 
    & \textbf{[C2]}  Perform network operations with networking libraries such as requests and urllib for HTTP requests and responses\\
    & \textbf{[C3]} Implement server and client programming, including creating TCP/UDP servers and clients for communication\\ 
    & \textbf{[C4]} Work with HTTP protocols, handling methods like GET, POST, PUT, and DELETE, as well as headers, status codes, and responses\\ 
    \midrule

    \textbf{[K19]} Serialization & 
    \textbf{[C1]}  Perform JSON serialization and deserialization using the json module (json.dump(), json.dumps(), json.load(), json.loads())\\ 
    & \textbf{[C2]}  Store and retrieve Python objects with pickling using the pickle module (pickle.dump(), pickle.load())\\
    & \textbf{[C3]} Parse and process XML data using modules such as xml.etree.ElementTree for reading, writing, and navigating XML structures\\ 
    \midrule

    \textbf{[K20]} Database & 
    \textbf{[C1]}  Execute core SQL queries: SELECT, INSERT, UPDATE, DELETE\\ 
    & \textbf{[C2]}  Apply filtering/sorting/limits with WHERE, ORDER BY, LIMIT/OFFSET, and patterns (LIKE, IN, BETWEEN) and aggregation (COUNT, SUM, AVG, MIN, MAX with GROUP BY and HAVING)\\
    & \textbf{[C3]} Perform joins: INNER, LEFT (and others as supported), including multi-table joins\\ 
    & \textbf{[C4]} Define and evolve schema: CREATE TABLE, ALTER, constraints (PRIMARY/FOREIGN KEY, UNIQUE), and indexes\\ 
    & \textbf{[C5]} Use ORMs (e.g., SQLAlchemy): define models, sessions, CRUD operations, relationships, and query building\\ 
    \midrule
    
    \end{tabular}
    }
\end{table}



\FloatBarrier
\section{Examples of generated KU-based tasks} \label{appendix:ku_baesd_task_examples}

\begin{figure}[!h]
    \centering
    \includegraphics[width=1.0\textwidth]{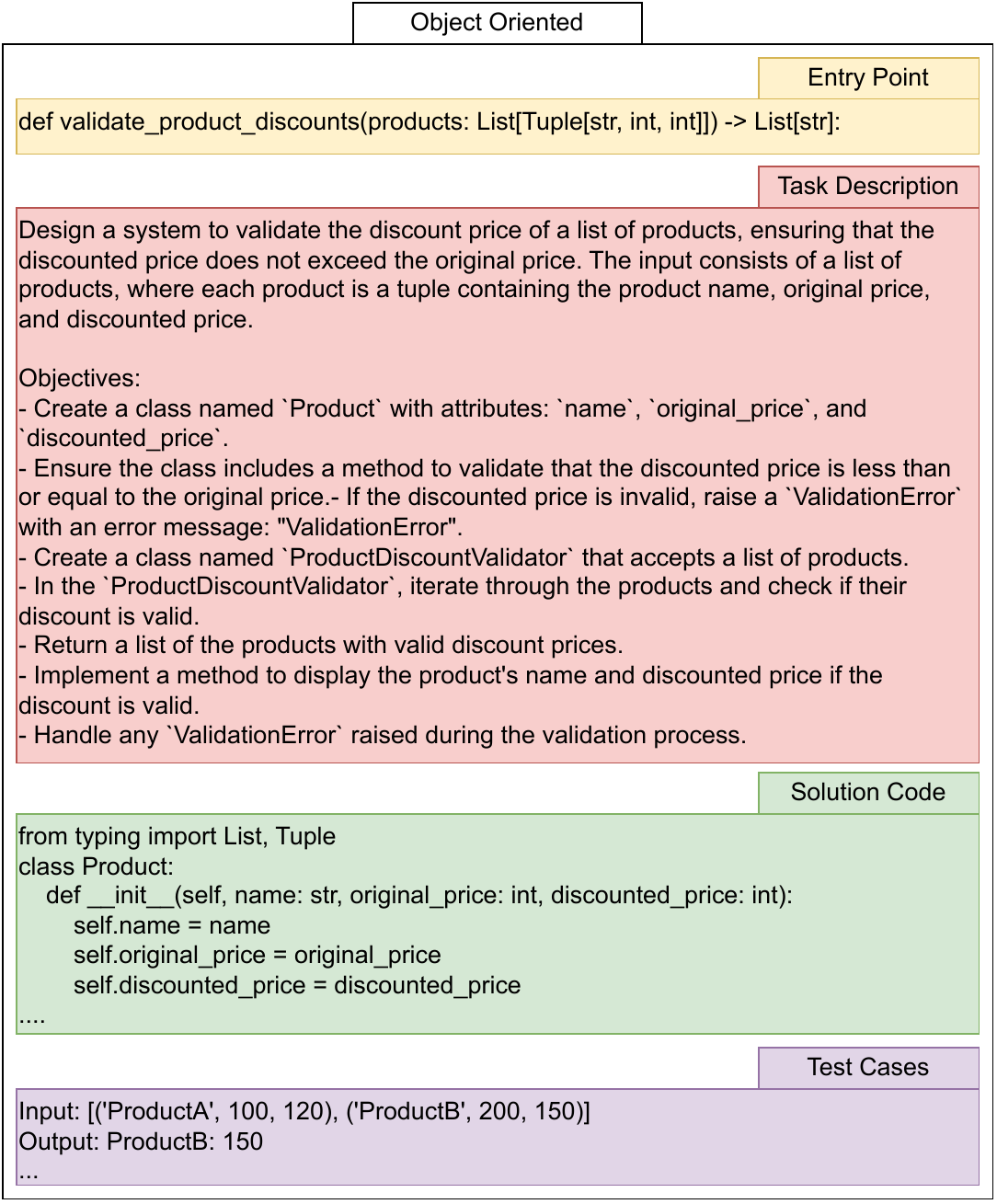}
    \caption{Example of a generated KU-based task for the Object Oriented Programming KU.}
    \label{fig:task_1}
\end{figure}

\begin{figure}[!h]
    \centering
    \includegraphics[width=1.0\textwidth]{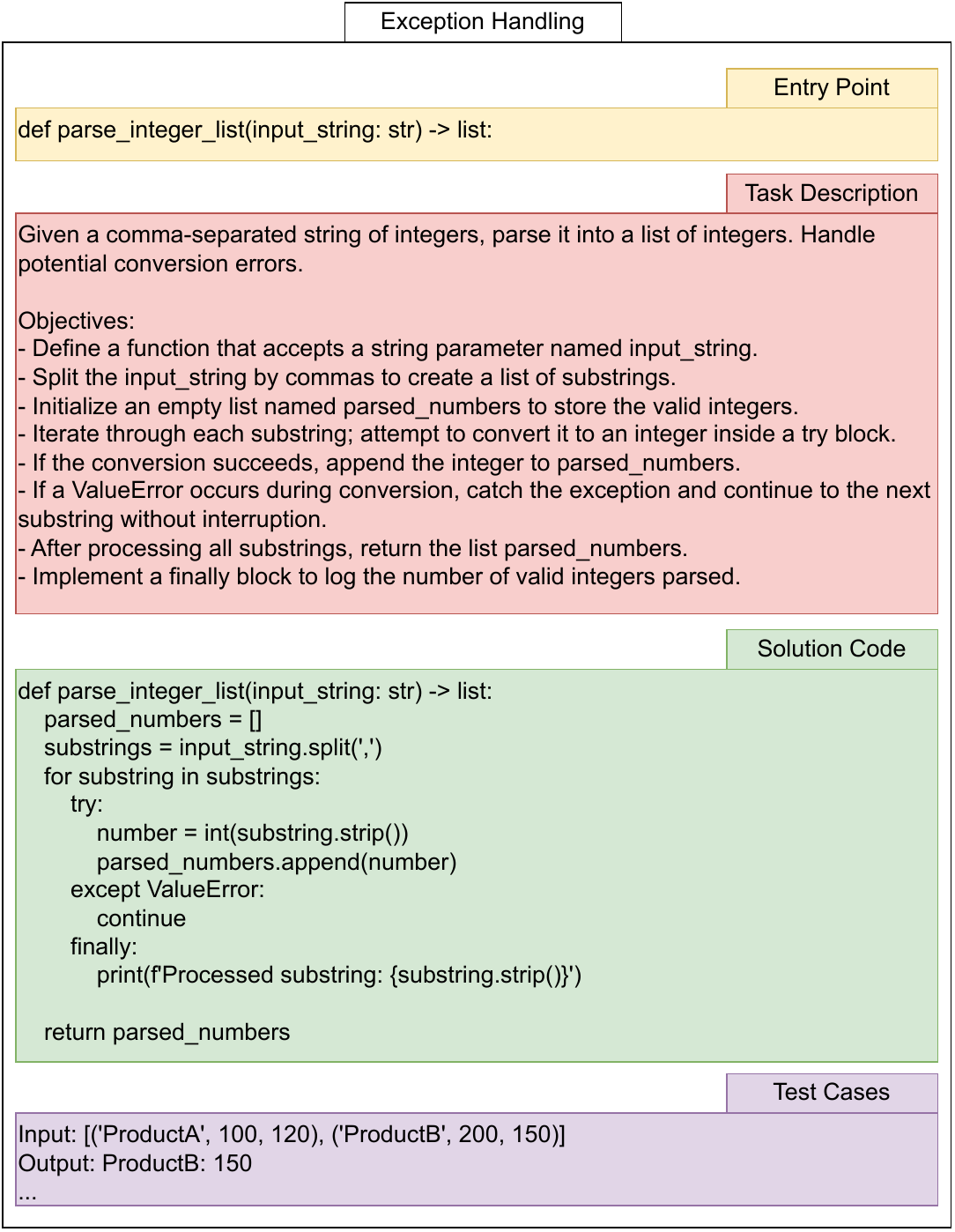}
    \caption{Example of a generated KU-based tasks for the Exception Handling KU.}
    \label{fig:task_2}
\end{figure}

\begin{figure}[!h]
    \centering
    \includegraphics[width=1.0\textwidth]{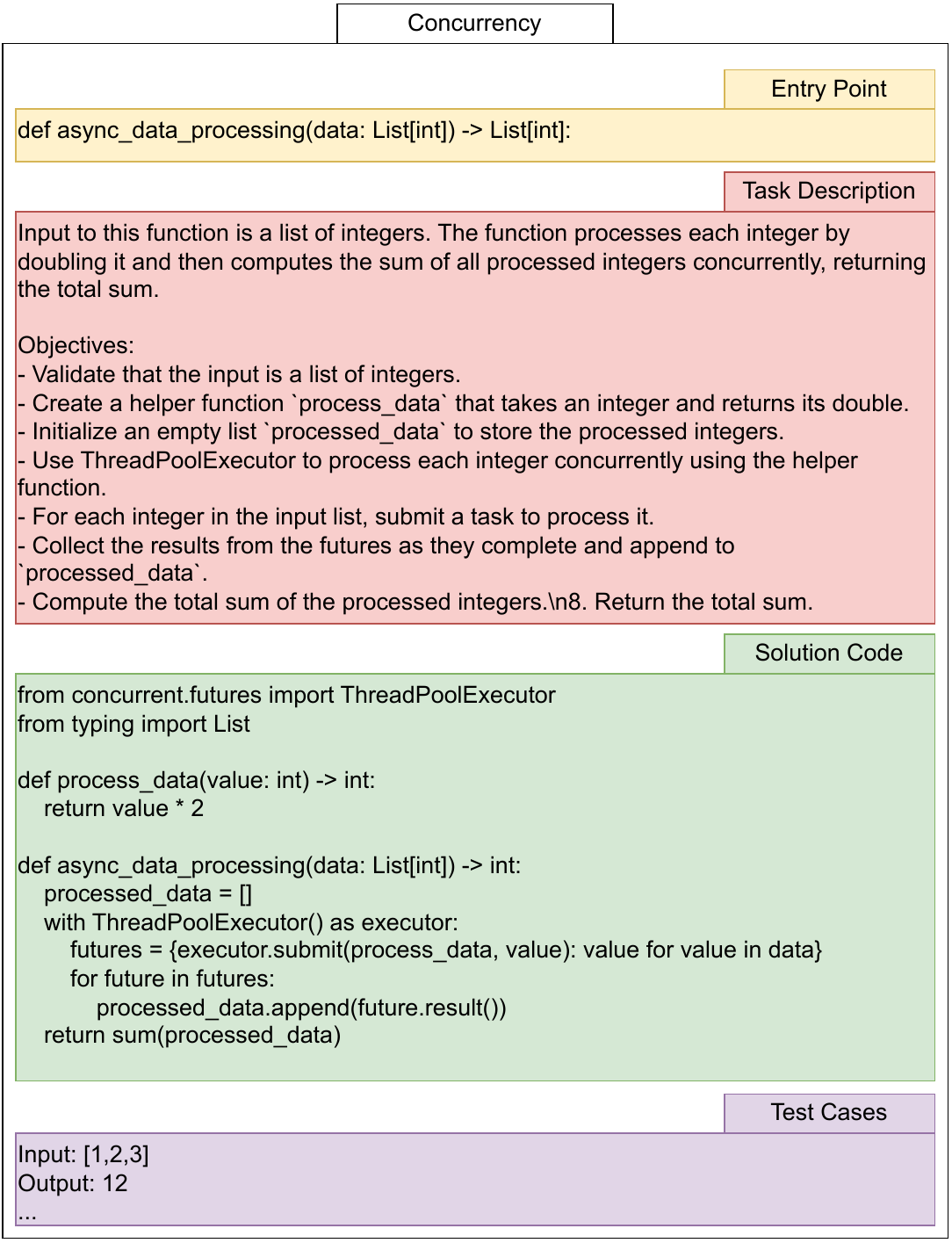}
    \caption{Example of a generated KU-based task for the Concurrency KU.}
    \label{fig:task_3}
\end{figure}

\FloatBarrier
\section{Comparison results of the performance different LLMs for Augmented-MBPP} \label{appendix:ku_augment_MBPP}

\begin{table}[!h]
\centering
\small
\caption{Comparison of the performance of different LLMs across MBPP, newly generated KU-based tasks, and Augmented-MBPP. The last row for each model reports the {\color{red}relative performance drop (percentage)} from MBPP to Augmented-MBPP.}
\label{tab:model-performance-mbpp}
\begin{tabular}{@{}llrrr@{}}
\toprule
\multicolumn{1}{c}{\textbf{Model}} & \multicolumn{1}{c}{\textbf{Task\_Type}}                                                                            & \textbf{Pass@1}              & \textbf{Pass@3}              & \textbf{Pass@5}              \\ \midrule
                                   & MBPP                                                                                                               & 0.59                         & 0.74                         & 0.78                         \\
                                   & NewKUTasks                                                                                                         & 0.49                         & 0.55                         & 0.57                         \\
                                   & Augmented-MBPP                                                                                                          & 0.57                         & 0.70                         & 0.73                         \\  \cmidrule(l){2-5} 
\multirow{-3}{*}{LLaMa3}           & {\color[HTML]{FE0000} \begin{tabular}[c]{@{}l@{}}Relative Performance Drop\\ (HumanEval → Augmented)\end{tabular}} & {\color[HTML]{FE0000} 4.02}  & {\color[HTML]{FE0000} 5.81}  & {\color[HTML]{FE0000} 6.11}  \\ \midrule
                                   & MBPP                                                                                                               & 0.54                         & 0.68                         & 0.72                         \\
                                   & NewKUTasks                                                                                                         & 0.43                         & 0.54                         & 0.58                         \\
                                   & Augmented-MBPP                                                                                                     & 0.51                         & 0.65                         & 0.69                         \\  \cmidrule(l){2-5} 
\multirow{-4}{*}{StarCoder2}       & {\color[HTML]{FE0000} \begin{tabular}[c]{@{}l@{}}Relative Performance Drop\\ (HumanEval → Augmented)\end{tabular}} & {\color[HTML]{FE0000} 4.65}  & {\color[HTML]{FE0000} 4.70}  & {\color[HTML]{FE0000} 4.56}  \\ \midrule
                                   & MBPP                                                                                                               & 0.53                         & 0.67                         & 0.72                         \\
                                   & NewKUTasks                                                                                                         & 0.41                         & 0.51                         & 0.55                         \\
                                   & Augmented-MBPP                                                                                                     & 0.50                         & 0.63                         & 0.68                         \\
\multirow{-4}{*}{Granite}          & {\color[HTML]{FE0000} \begin{tabular}[c]{@{}l@{}}Relative Performance Drop\\ (HumanEval → Augmented)\end{tabular}} & {\color[HTML]{FE0000} 5.03}  & {\color[HTML]{FE0000} 5.31}  & {\color[HTML]{FE0000} 5.36}  \\ \midrule
                                   & MBPP                                                                                                               & 0.38                         & 0.54                         & 0.60                         \\
                                   & NewKUTasks                                                                                                         & 0.18                         & 0.25                         & 0.29                         \\
                                   & Augmented-MBPP                                                                                                     & 0.33                         & 0.47                         & 0.53                         \\  \cmidrule(l){2-5} 
\multirow{-4}{*}{Mixtral}          & {\color[HTML]{FE0000} \begin{tabular}[c]{@{}l@{}}Relative Performance Drop\\ (HumanEval → Augmented)\end{tabular}} & {\color[HTML]{FE0000} 12.23} & {\color[HTML]{FE0000} 12.29} & {\color[HTML]{FE0000} 11.91} \\ \midrule
                                   & MBPP                                                                                                               & 0.72                         & 0.75                         & 0.76                         \\
                                   & NewKUTasks                                                                                                         & 0.53                         & 0.55                         & 0.57                         \\
                                   & Augmented-MBPP                                                                                                     & 0.67                         & 0.70                         & 0.71                         \\ \cmidrule(l){2-5} 
\multirow{-4}{*}{Gemma3}           & {\color[HTML]{FE0000} \begin{tabular}[c]{@{}l@{}}Relative Performance Drop\\ (HumanEval → Augmented)\end{tabular}} & {\color[HTML]{FE0000} 6.07}  & {\color[HTML]{FE0000} 5.99}  & {\color[HTML]{FE0000} 5.68}  \\ \midrule
                                   & MBPP                                                                                                               & 0.63                         & 0.68                         & 0.69                         \\
                                   & NewKUTasks                                                                                                         & 0.16                         & 0.19                         & 0.20                         \\
                                   & Augmented-MBPP                                                                                                     & 0.53                         & 0.57                         & 0.59                         \\ \cmidrule(l){2-5} 
\multirow{-4}{*}{GPT-3.5-Turbo}    & {\color[HTML]{FE0000} \begin{tabular}[c]{@{}l@{}}Relative Performance Drop\\ (HumanEval → Augmented)\end{tabular}} & {\color[HTML]{FE0000} 15.61} & {\color[HTML]{FE0000} 15.25} & {\color[HTML]{FE0000} 14.96} \\ \midrule
                                   & MBPP                                                                                                               & 0.72                         & 0.73                         & 0.74                         \\
                                   & NewKUTasks                                                                                                         & 0.46                         & 0.54                         & 0.57                         \\
                                   & Augmented-MBPP                                                                                                     & 0.66                         & 0.69                         & 0.70                         \\ \cmidrule(l){2-5} 
\multirow{-4}{*}{GPT-4o-Mini}      & {\color[HTML]{FE0000} \begin{tabular}[c]{@{}l@{}}Relative Performance Drop\\ (HumanEval → Augmented)\end{tabular}} & {\color[HTML]{FE0000} 7.67}  & {\color[HTML]{FE0000} 5.49}  & {\color[HTML]{FE0000} 4.89}  \\ \bottomrule
\end{tabular}
\end{table}

\end{document}